\documentclass[iop]{emulateapj}
\newcommand{\LX}{\ensuremath{L_{\mathrm{X}}}}

\newcommand{\Ne}{\ensuremath{n_{\mathrm{e}}}}

\newcommand{\NH}{\ensuremath{N_{\mathrm{H}}}}

\newcommand{\Rsol}{\ensuremath{R_{\odot}}}

\newcommand{\vc}{\ensuremath{v_\mathrm{c}}}

\newcommand{\nm}{\ensuremath{\mbox{\nm}}}
\newcommand{\cm}{\ensuremath{\mbox{cm}}}

\newcommand{\km}{\ensuremath{\mbox{km}}}

\newcommand{\pc}{\ensuremath{\mbox{pc}}}
\newcommand{\kpc}{\ensuremath{\mbox{kpc}}}

\newcommand{\s}{\ensuremath{\mbox{s}}}
\newcommand{\ks}{\ensuremath{\mbox{ks}}}

\newcommand{\yr}{\ensuremath{\mbox{yr}}}
\newcommand{\Myr}{\ensuremath{\mbox{Myr}}}

\newcommand{\kev}{\ensuremath{\mbox{keV}}}

\newcommand{\erg}{\ensuremath{\mbox{erg}}}

\newcommand{\sr}{\ensuremath{\mbox{sr}}}

\newcommand{\microgauss}{\ensuremath{\mu\mbox{G}}}

\newcommand{\K}{\ensuremath{\mbox{K}}}

\newcommand{\ph}{\ensuremath{\mbox{photons}}}
\newcommand{\counts}{\ensuremath{\mbox{counts}}}

\newcommand{\parcminsq}{\ensuremath{\mbox{arcmin}^{-2}}}
\newcommand{\pdegsq}{\ensuremath{\mbox{deg}^{-2}}}
\newcommand{\pcc}{\ensuremath{\cm^{-3}}}
\newcommand{\pcmsq}{\ensuremath{\cm^{-2}}}

\newcommand{\ps}{\ensuremath{\s^{-1}}}
\newcommand{\psr}{\ensuremath{\sr^{-1}}}

\newcommand{\emismeas}{\ensuremath{\cm^{-6}}\ \pc}
\newcommand{\ergps}{\erg\ \ps}
\newcommand{\flux}{\erg\ \pcmsq\ \ps}

\newcommand{\kmps}{\km\ \ps}

\newcommand{\pownorm}{\ph\ \pcmsq\ \ps\ \psr\ \ensuremath{\kev^{-1}}}

\newcommand{\presalt}{\pcc\ \K}
\newcommand{\rassrate}{\counts\ \ps\ \parcminsq}

\newcommand{\HI}{H~\textsc{i}}

\newcommand{\OVII}{O~\textsc{vii}}
\newcommand{\OVIII}{O~\textsc{viii}}

\newcommand{\chandra}{\textit{Chandra}}

\newcommand{\rosat}{\textit{ROSAT}}

\newcommand{\suzaku}{\textit{Suzaku}}

\newcommand{\xmm}{\textit{XMM-Newton}}

\newcommand{\citepossessive}[1]{\citeauthor{#1}'s \citeyearpar{#1}}

\newcommand{\e}{\ensuremath{\mathrm{e}}}
\newcommand{\chisq}{\ensuremath{\chi^2}}

\newcommand{\raymondsmith}{\citeauthor{raymond77} (\citeyear{raymond77} and updates)}
\newcommand{\eqref}[1]{equation~(\ref{#1})}

\newcommand{\Beff}{\ensuremath{B_\mathrm{eff}}}

\newcommand{\EMsinb}{\ensuremath{\mathrm{E.M.} \sin |b|}}
\newcommand{\hSPH}{\ensuremath{h_\mathrm{SPH}}}
\newcommand{\Pnt}{\ensuremath{P_\mathrm{nt}}}
\newcommand{\Rcyl}{\ensuremath{R_\mathrm{cyl}}}
\newcommand{\Rsph}{\ensuremath{R_\mathrm{sph}}}

\newcommand{\Stotal}{\ensuremath{S_{0.4-2.0}}}
\newcommand{\SX}{\ensuremath{S_\mathrm{X}}}

\shorttitle{HOT GAS IN THE GALACTIC HALO}
\shortauthors{HENLEY ET AL.}

\begin{document}

\title{The Origin of the Hot Gas in the Galactic Halo: Confronting Models with \textit{XMM-Newton} Observations}
\author{David B. Henley\altaffilmark{1},
        Robin L. Shelton\altaffilmark{1},
        Kyujin Kwak\altaffilmark{1},
        M. Ryan Joung\altaffilmark{2,3}, and
        Mordecai-Mark Mac Low\altaffilmark{3}
}
\altaffiltext{1}{Department of Physics and Astronomy, University of Georgia, Athens, GA 30602}
\altaffiltext{2}{Department of Astronomy, Columbia University, 550 West 120th Street, New York, NY 10027}
\altaffiltext{3}{Department of Astrophysics, American Museum of Natural History, 79th Street at Central Park West, New York, NY 10024}
\email{dbh@physast.uga.edu}

\begin{abstract}
We compare the predictions of three physical models for the origin of the hot halo gas with the
observed halo X-ray emission, derived from 26 high-latitude \xmm\ observations of the soft X-ray
background between $l=120\degr$ and $l=240\degr$. These observations were chosen from a much larger
set of observations as they are expected to be the least contaminated by solar wind charge exchange
emission. We characterize the halo emission in the \xmm\ band with a single-temperature plasma model.
We find that the observed halo temperature is fairly constant across the sky
($\sim$$(\mbox{1.8--2.4}) \times 10^6~\K$), whereas the halo emission measure varies by an order of
magnitude ($\sim$0.0005--0.006~\emismeas).
When we compare our observations with the model predictions, we find that most of the hot gas
observed with \xmm\ does not reside in isolated extraplanar supernova remnants -- this model predicts emission an
order of magnitude too faint. A model of a supernova-driven interstellar medium, including the flow of
hot gas from the disk into the halo in a galactic fountain, gives good agreement with the observed
0.4--2.0~\kev\ surface brightness. This model overpredicts the halo X-ray temperature by a factor of
$\sim$2, but there are a several possible explanations for this discrepancy. We therefore conclude
that a major (possibly dominant) contributor to the halo X-ray emission observed with \xmm\ is a
fountain of hot gas driven into the halo by disk supernovae.  However, we cannot rule out the
possibility that the extended hot halo of accreted material predicted by disk galaxy formation
models also contributes to the emission.
\end{abstract}


\section{INTRODUCTION}
\label{sec:Introduction}

Observations of the diffuse soft X-ray background (SXRB) indicate the presence of
$\sim$$(1\mbox{--}3) \times 10^6~\K$ X-ray-emitting gas in the interstellar medium (ISM) of our Galaxy.
Early observations with rocket-borne instruments led to the conclusion that the diffuse
1/4-\kev\ emission was dominated by emission from $\sim$$1 \times 10^6$~\K\ plasma in
the Local Bubble (LB), a $\sim$100-pc cavity in the ISM in which the Solar System resides
\citep{sanders77,snowden90}. The discovery of shadows in the 1/4-\kev\ background with
\rosat\ showed that there was also gas with $T \sim 1 \times 10^6~\K$ beyond the Galactic
disk, in the Galactic halo \citep{burrows91,snowden91}. Higher-energy emission data from \rosat,
\xmm, and \suzaku, and X-ray absorption line data from \chandra, show the presence of hotter gas in
the Galactic halo, with temperatures up to $\sim$$3 \times 10^6~\K$
\citep{kuntz00,yao05,yao07a,smith07a,galeazzi07,henley08a,yao09,lei09,yoshino09}.

Several possible sources for the hot halo gas have been suggested, including supernova- (SN) and
stellar wind-driven outflows from the Galactic disk \citep[e.g.,][]{shapiro76,bregman80,norman89},
gravitational heating of infalling intergalactic material (predicted by simulations of disk galaxy
formation; \citealp{toft02,rasmussen09}), and \textit{in situ} heating by extraplanar SNe
\citep{shelton06,henley09}.  X-ray spectroscopy is essential for determining which process or
processes have produced the $\sim$$(1\mbox{--}3) \times 10^6~\K$ gas in the halo. In principle, the observed
ionization state could be used to distinguish between the different models. For
example, gas heated by SNe could be underionized if heated recently, or overionized if heated in the
distant past \citep[e.g.,][]{shelton99}, and gas that has recently burst out of the disk, cooling
rapidly, will be drastically overionized \citep{breitschwerdt94}. The elemental abundance ratios
could also, in principle, be used to distinguish between models, as the abundance pattern
of the hot gas may depend on whether it is of Galactic or extragalactic origin.

In practice, it is not easy to use arguments based on the ionization state or the abundances to
distinguish between models, as collisional ionization equilibrium (CIE) models with solar abundances
generally provide good fits to the observed X-ray spectra
\citep[e.g.][]{galeazzi07,henley08a,lei09,yoshino09}, although supersolar [Ne/O] and [Fe/O]
abundance ratios have been reported for some sightlines \citep{yoshino09,yao09}.
Here, we use a different
approach. We fit CIE models to 26 \xmm\ spectra of the SXRB, obtained from observations between
$l=120\degr$ and $l=240\degr$ and with $|b| > 30\degr$. These fits yield temperatures and emission
measures for the halo. We then compare the measured distributions of these quantities to those
predicted by two physical models of the hot halo gas: a model in which the hot gas is heated
\textit{in situ} by extraplanar SNe and is contained in isolated supernova remnants (SNRs;
\citealt{shelton06}), and a model of an SN-driven ISM, one feature of which is the transfer of hot gas
from the disk to the halo \citep{joung06}. In addition, we use our observed halo parameters
to estimate the X-ray luminosity of the halo, and compare it to the predictions of disk galaxy
formation models \citep{toft02,rasmussen09,crain10}.

Our \xmm\ observations are a subset of those used in the survey of \citet[hereafter
  Paper~I]{henley10a}, who measured the SXRB \OVII\ and \OVIII\ intensities from 590 archival
\xmm\ observations between $l=120\degr$ and $l=240\degr$. The observations used here were chosen
because they should be less affected by solar wind charge exchange (SWCX) emission
\citep{cravens00}, which is a time-varying contaminant of SXRB spectra
\citep{cravens01,snowden04,fujimoto07,koutroumpa07,kuntz08a,carter08,henley08a}.  Although
the 26 \xmm\ observations used here are only a small subset of the observations used in Paper~I,
this is a larger number of observations than has been used in previous studies of the SXRB and the
hot ISM with CCD-resolution spectra

The remainder of this paper is organized as follows. In Section~\ref{sec:Observations} we present
the details of our observations and give an overview of the data reduction (see Paper~I for more
details).  Section~\ref{sec:SpectralAnalysis} contains our spectral analysis, in which we use CIE
models to determine the spectrum of the halo emission. In Section~\ref{sec:ComparisonWithModels} we
compare the results of our spectral analysis with the predictions of various physical models for the
origin of the hot halo gas. In particular, the disk galaxy formation model, the extraplanar SN
model, and the SN-driven ISM model are presented in Sections~\ref{subsec:DiskGalaxyFormation},
\ref{subsec:InSituSupernovae}, and \ref{subsec:SNDrivenISM}, respectively. We discuss our results in
Section~\ref{sec:Discussion}, and conclude with a summary in Section~\ref{sec:Summary}.  Throughout
we use \citet{anders89} abundances.

\section{OBSERVATIONS}
\label{sec:Observations}

Our sample of observations is taken from Paper~I, which presents \OVII\ and \OVIII\ intensities
extracted from 590 \xmm\ observations between $l = 120\degr$ and $l = 240\degr$. The observations
used here were selected by applying various filters to minimize the contamination from SWCX
emission. In particular, to minimize geocoronal and near-Earth heliospheric SWCX contamination, we
rejected the portions of our \xmm\ data taken when the solar wind proton flux\footnote{The solar
  wind proton flux data were obtained from OMNIWeb (http://omniweb.gsfc.nasa.gov).} exceeded $2 \times
10^8~\pcmsq~\ps$, and to minimize heliospheric SWCX contamination we used only observations of high
ecliptic latitudes ($\beta > 20\degr$) taken during solar minimum.\footnote{Following Paper~I, we used
  observations made after 00:00UT on 2005 Jan 01. This data was estimated using sunspot data from
  the National Geophysical Data Center (http://www.ngdc.noaa.gov/stp/SOLAR/ftpsunspotnumber.html).}
See Section~2 of Paper~I for more details about SWCX and the filters we used to reduce its effects.

\begin{figure}
\centering
\plotone{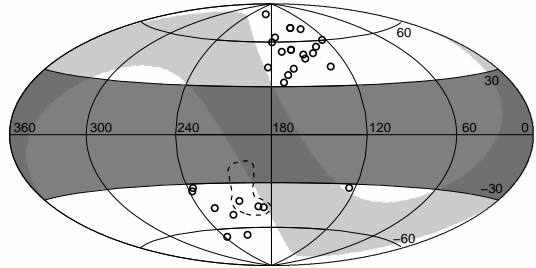}
\caption{All-sky Hammer-Aitoff projection in Galactic coordinates, centered on the Galactic
  Anticenter, showing the \xmm\ pointing directions used in this study. The gray bands indicate
  regions of the sky that were excluded from our sample -- the darker band shows $|b| \le 30\degr$,
  and the lighter band shows $|\beta| \le 20\degr$. The dashed line outlines the Eridanus Enhancement.
  The three observations toward this feature were also excluded.
  \label{fig:ObservationLocations}}
\end{figure}

After applying these filters, 43 observations remained (see Section~5.3.1 of Paper~I).
As we are interested in the Galactic halo here, we removed 14 more observations at low Galactic
latitudes ($|b| \le 30\degr$). The locations of the 29 remaining observations on the sky are shown
in Figure~\ref{fig:ObservationLocations}. Three of these observations are toward the Eridanus
Enhancement, a large, X-ray--bright superbubble \citep{burrows93,snowden95b}. We also removed these
3 observations. The details of the 26 observations that remain in our sample are summarized in
Table~\ref{tab:ObservationDetails}.

\begin{deluxetable*}{rllrrcccc}
\tabletypesize{\small}
\tablewidth{0pt}
\tablecaption{\xmm\ Observation Details\label{tab:ObservationDetails}}
\tablehead{
& 			&			&			&			& \multicolumn{2}{c}{MOS1}			& \multicolumn{2}{c}{MOS2}				\\
& \colhead{Obs. Id}	& \colhead{Start Date}	& \colhead{$l$}		& \colhead{$b$}		& \colhead{Exposure}	& \colhead{$\Omega$}	& \colhead{Exposure}	& \colhead{$\Omega$}		\\
&			&			& \colhead{(deg)}	& \colhead{(deg)}	& \colhead{(ks)}	& \colhead{(arcmin$^2$)}& \colhead{(ks)}	& \colhead{(arcmin$^2$)}	\\
& \colhead{(1)}		& \colhead{(2)}		& \colhead{(3)}		& \colhead{(4)}		& \colhead{(5)}		& \colhead{(6)}		& \colhead{(7)}		& \colhead{(8)}			\\
}
\startdata
 1 & 0304070501 & 2005-11-08 & 124.223 & $ 60.304$ &            12.0 &             486 &            12.0 &             510 \\
 2 & 0305290201 & 2005-07-02 & 124.578 & $-32.485$ &            15.1 &             478 &            16.9 &             572 \\
 3 & 0401210601 & 2006-10-10 & 133.225 & $ 42.419$ &            17.7 &             310 &            17.6 &             394 \\
 4 & 0404220101 & 2006-11-01 & 135.974 & $ 55.981$ &            13.0 &             488 &            14.1 &             505 \\
 5 & 0400560301 & 2006-11-17 & 138.279 & $ 68.853$ &            51.5 &             380 &            51.5 &             403 \\
 6 & 0400570201 & 2006-11-25 & 142.370 & $ 51.705$ &            23.0 &             463 &            22.9 &             475 \\
 7 & 0406630201 & 2007-04-12 & 151.186 & $ 48.245$ &             8.5 &             339 &             8.4 &             431 \\
 8 & 0303260201 & 2005-04-07 & 151.607 & $ 51.006$ &            44.4 &             411 &            44.0 &             583 \\
 9 & 0306060201 & 2005-11-13 & 151.829 & $ 70.103$ &            53.5 &             410 &            54.8 &             511 \\
10 & 0306060301 & 2005-11-15 & 151.831 & $ 70.103$ &            15.3 &             415 &            15.6 &             511 \\
11 & 0303720601 & 2005-04-25 & 161.440 & $ 54.439$ &            23.5 &             382 &            22.9 &             398 \\
12 & 0303720201 & 2005-04-13 & 161.441 & $ 54.439$ &            25.8 &             378 &            26.2 &             470 \\
13 & 0200960101 & 2005-03-28 & 162.721 & $ 41.656$ &            56.9 &             453 &            57.0 &             465 \\
14 & 0301340101 & 2006-04-12 & 167.648 & $ 37.517$ &            12.8 &             487 &            13.0 &             512 \\
15 & 0306370601 & 2005-04-24 & 170.477 & $ 53.178$ &             9.6 &             496 &            10.1 &             583 \\
16 & 0402780701 & 2007-03-28 & 171.132 & $ 32.731$ &            14.3 &             422 &            14.6 &             577 \\
17 & 0304203401 & 2006-06-11 & 175.807 & $ 63.353$ &             8.5 &             371 &             8.3 &             391 \\
18 & 0406610101 & 2006-11-05 & 179.356 & $ 59.942$ &            10.4 &             402 &            10.8 &             485 \\
19 & 0400830301 & 2006-10-30 & 182.658 & $ 42.566$ &            45.0 &             493 &            44.7 &             511 \\
20 & 0301651701 & 2006-06-20 & 197.309 & $ 81.121$ &            12.3 &             473 &            12.2 &             495 \\
21 & 0300630301 & 2006-01-19 & 209.821 & $-65.146$ &            14.4 &             477 &            14.5 &             499 \\
22 & 0312190601 & 2006-01-28 & 213.849 & $-50.846$ &            11.3 &             391 &            11.2 &             477 \\
23 & 0301330401 & 2006-02-13 & 226.946 & $-45.906$ &            19.5 &             414 &            19.6 &             583 \\
24 & 0312190701 & 2006-01-28 & 236.040 & $-32.583$ &            11.1 &             483 &            10.8 &             571 \\
25 & 0302500101 & 2005-08-09 & 237.074 & $-65.638$ &            21.9 &             413 &            23.5 &             583 \\
26 & 0307001401 & 2006-02-13 & 237.615 & $-34.679$ &             7.8 &             489 &             8.3 &             584 \\

\enddata
\tablecomments{The observations are in order of increasing $l$.
Column~1 contains the \xmm\ observation ID.
Column~2 contains the observation start date, in YYYY-MM-DD format.
Columns~3 and 4 contain the pointing direction in Galactic coordinates.
Column~5 contains the usable MOS1 exposure time that remains after the filtering mentioned in Section~\ref{sec:Observations}.
Column~6 contains the solid angle, $\Omega$, from which the MOS1 SXRB spectrum was extracted, after the removal of sources and unusable CCDs.
Columns~7 and 8 contain the corresponding data for MOS2.}
\end{deluxetable*}

The data reduction was carried out using SAS version
7.0.0\footnote{http://xmm2.esac.esa.int/sas/7.0.0/} and the \xmm\ Extended Source Analysis
Software\footnote{http://heasarc.gsfc.nasa.gov/docs/xmm/xmmhp\_xmmesas.html} (XMM-ESAS) version 2
\citep{snowden07,kuntz08a}. We only used data from the EPIC-MOS cameras \citep{turner01}, as the
version of XMM-ESAS that we used cannot calculate the particle background for EPIC-pn data
\citep{struder01}. The data reduction method is described in full in Paper~I. Here we outline the
main steps.

We cleaned and filtered each dataset using the XMM-ESAS \texttt{mos-filter} script. This script runs
the standard \texttt{emchain} processing script, and then uses the XMM-ESAS \texttt{clean-rel}
program to identify and remove times affected by soft-proton flaring. As noted above, we also
removed times when the solar wind proton flux exceeded $2 \times 10^8~\pcmsq~\ps$. The usable
exposures for the two MOS cameras are shown in Columns~5 and 7 of
Table~\ref{tab:ObservationDetails}.

We detected and removed point sources with a 0.5--2.0~\kev\ flux greater than $5 \times
10^{-14}~\flux$.  This is the approximate flux level to which \citet{chen97} removed sources when
they measured the spectrum of the extragalactic background to be $10.5 (E /
1~\kev)^{-1.46}~\pownorm$ (we use this model in our spectral analysis; see
Section~\ref{subsec:ModelDescription}). Some observations contained sources that were too bright or
too extended to be removed by the automated source removal. We removed such sources by hand, by
excluding a circular region centered on each source.

We extracted SXRB spectra from the entire field of view, minus any sources that were removed, and
minus any CCDs that were in window mode or that exhibited the anomalous state described by
\citet{kuntz08a}. The solid angles from which each spectrum was extracted are shown in columns~6 and
8 of Table~\ref{tab:ObservationDetails}. We binned the spectra such that each bin contained at least
25 counts.  We created redistribution matrix files (RMFs) and ancillary response files (ARFs) using
\texttt{rmfgen} and \texttt{arfgen}.

We used the XMM-ESAS \texttt{xmm-back} program to calculate the quiescent particle background (QPB)
spectrum for each observation. The QPB spectra were constructed from a database of
filter-wheel-closed data, scaled using data from the unexposed corner pixels that lie outside the
field of view (see \citealt{kuntz08a} for more details of the modeling of the QPB spectrum). The QPB
spectra were subtracted from the corresponding SXRB spectra before we carried out our spectral
analysis.

\section{SPECTRAL ANALYSIS}
\label{sec:SpectralAnalysis}

\subsection{Model Description}
\label{subsec:ModelDescription}

We analyzed the spectra from each of our 26 \xmm\ observations using a multicomponent model of the SXRB.
The model consisted of the following components: a foreground emission component (representing LB
emission and/or SWCX emission that remains in our spectra, despite our efforts to minimize this contamination),
a Galactic halo component, an extragalactic component, and instrumental background components.

We used a \raymondsmith\ model with $T \sim 10^6~\K$ to model the foreground emission in our
analysis, because such a model provides a good fit to data from the \rosat\ All-Sky Survey
\citep[e.g.,][]{snowden98,snowden00,kuntz00}. We fixed the temperature ($T = 1.2 \times 10^6~\K$) and
normalization of this component using data from the \citet{snowden00} catalog of shadows in the
\rosat\ All-Sky Survey. Following Paper~I, we calculated the foreground R12 (1/4~\kev) count-rate
for each \xmm\ sightline by averaging the foreground count-rates of the 5 nearest shadows, weighted
by their inverse distances from the \xmm\ sightline, i.e.,
\begin{equation}
\mbox{Foreground R12 count-rate} = \frac{\sum R_i / \theta_i}{\sum 1 / \theta_i},
\end{equation}
where $R_i$ is the foreground R12 count-rate for the $i$th shadow, and $\theta_i$ is the angle
between the center of the $i$th shadow and the \xmm\ sightline. Over the whole set of
\xmm\ sightlines, the median value of $\theta_i$ is 6.2\degr\ (lower and upper quartiles:
4.1\degr\ and 7.7\degr, respectively).  The median minimum and maximum values of $\theta_i$ for each
sightline are 2.8\degr\ and 7.4\degr, respectively. Only three sightlines have $\theta_i > 7\degr$
for all $i$: obs.~0305290201 ($\theta_i = 7.4\degr$--9.3\degr), obs.~0301340101 ($\theta_i =
8.6\degr$--14.5\degr), and obs.~0402780701 ($\theta_i = 11.2\degr$--14.7\degr).
We used the above-calculated count-rate to determine the normalization of the foreground component
for the sightline in question (assuming $T = 1.2 \times 10^6~\K$). This normalization was held fixed
during the subsequent spectral fitting.

For the Galactic halo, which is the component of the SXRB that we are interested in here, we used a
single-temperature ($1T$) \raymondsmith\ model. While such a halo model is inadequate for explaining
all the available X-ray and far-ultraviolet data \citep{yao07a,shelton07,lei09}, it is adequate for
characterizing the X-ray emission in the \xmm\ band. The temperature and emission measure of this
component were both free parameters in the spectral fitting.  We used a \citeauthor{raymond77}
model, instead of, say, an APEC model, because the codes that we used to calculate X-ray spectra
from hydrodynamical models (see Section~\ref{sec:ComparisonWithModels}) also use the
\citeauthor{raymond77} code.  In our analysis, the temperature and emission measure of the Galactic
halo component were free to vary.

We modeled the extragalactic background as a power-law with a photon index $\Gamma = 1.46$ and a normalization of 10.5~\pownorm\
at 1~\kev\ \citep{chen97}. The extragalactic and halo components were both subject to absorption. For each sightline,
we used the HEAsoft \texttt{nh} tool to obtain the \HI\ column density from the Leiden-Argentine-Bonn (LAB) Survey of
Galactic \HI\ \citep{kalberla05}. We used photoelectric absorption cross-sections from \citet{balucinska92}, with an
updated He cross-section from \citet{yan98}.

As well as the above SXRB components, we included components to model parts of the particle
background. These components were independent for the two MOS detectors. The QPB spectrum includes
two bright fluorescent lines from aluminum and silicon at 1.49 and 1.74~\kev, respectively. The
procedure for calculating the QPB spectrum mentioned in Section~\ref{sec:Observations} cannot
adequately remove these lines. Instead, the QPB spectrum was interpolated between 1.2 and 1.9~\kev,
and we added two Gaussians to our spectral model to model these lines. In addition, despite the data
cleaning described in Section~\ref{sec:Observations}, some residual soft-proton contamination may
remain in the data. We modeled this contamination using a power-law which we did not fold through
the instrumental response \citep{snowden07,snowden10}. Following suggestions in the latest version
of the XMM-ESAS manual (\citealt{snowden10}, which pertains to a later version of the software than
the one we used), we placed constraints on the spectral index of this power-law (specifically, soft
limits at 0.5 and 1.0, and hard limits at 0.2 and 1.3).

Originally, as in Paper~I, we used a broken power-law to model the soft protons. We fixed the break
at 3.2~\kev\ \citep{kuntz08a}, but we did not impose any special constraints on the spectral
indices. We find that the halo temperatures are generally not significantly affected by our choice
of soft proton model, but the emission measures and surface brightnesses are typically 30\%\ higher
if we place constraints on the power-law spectral index. However, these differences do not affect
the conclusions of this paper. Throughout this paper, we use the newer set of spectral fit results,
obtained using an unbroken power-law with constraints on the spectral index to model the soft protons.

We carried out our spectral analysis using XSPEC\footnote{http://heasarc.gsfc.nasa.gov/docs/xanadu/xspec/} version 12.5.0.
We analyzed each of our 26 observations individually, fitting the above-described model to the 0.4--5.0~\kev\
MOS1 and MOS2 spectra simultaneously.

\subsection{Systematic Errors}
\label{subsec:Systematics}

Fixing the spectra of the foreground and extragalactic components may introduce systematic errors to
our fitting (i.e., the true spectra of these components may differ from our assumed spectra, which
may in turn affect the measured halo parameters). Here, we estimate the sizes of these systematic
errors.

In the case of the foreground model, we fixed the spectrum because otherwise there would be a
degeneracy between the foreground and background intensities. This degeneracy can be overcome by
shadowing observations \citep{smith07a,galeazzi07,henley08a,gupta09b,lei09}, but such an analysis is
not possible here. Here, we fixed the foreground normalization by extrapolating the foreground spectrum from
the R12 band into the \xmm\ band. However, as the relative contributions of LB and SWCX emission are
likely to differ in these bands, and these two emission mechanisms have different spectra, such an
extrapolation may lead to an incorrect estimate of the foreground normalization.  We estimated the
size of this systematic effect by reanalyzing each sightline, using the median foreground R12
intensity (600~\rassrate) to fix the normalization for every sightline. We used the median
differences between our original results and these new results as our estimate of the systematic errors due
to the foreground normalization being fixed (we adopted the same systematic errors for each sightline).
The estimated systematic errors are $\pm0.08 \times 10^6~\K$ for the halo temperature and
$\pm0.04$~dex for the halo emission measure.

We fixed the spectrum of the extragalactic background component because otherwise there would be a
degeneracy between this component and the power-law component used to model the soft protons.
However, the normalization of this component may vary from field to field, due to statistical
fluctuations in the number of unresolved sources that comprise the extragalactic background.  We
estimated the size of the field-to-field variation in the extragalactic background using the
0.5--2.0~\kev\ source flux distribution from \citet{moretti03}. Given that we removed sources with
fluxes greater than $5 \times 10^{-14}~\flux$, we estimate that the 0.5--2.0~\kev\ extragalactic
surface brightness varies by roughly $\pm10\%$ from field to field, assuming a typical field of view
of 480~arcmin$^2$. We therefore repeated our analysis for each sightline with assumed extragalactic
normalizations of 9.5 and 11.6~\pownorm. We used the differences between the original results and
these new results to estimate the systematic errors for each sightline due to our fixing the
normalization of the extragalactic background. The systematic errors on the halo temperature and
emission measure are typically less than $\pm0.2 \times 10^6~\K$ and $\pm0.2$~dex, respectively.

An additional possible source of systematic error is the expected steepening of the extragalactic
background below $\sim$2~\kev. In the 3--10~\kev\ energy range, the extragalactic photon index
$\Gamma \approx 1.4$ \citep{marshall80}. This photon index also provides good fits to SXRB spectra
at lower energies (e.g., \citealt{chen97}, the source of our assumed extragalactic background
model).  However, the summed spectrum of the individual sources that comprise the extragalactic
background is steeper below $\sim$2~\kev\ \citep[$\Gamma = 1.96$;][]{hasinger93}. Failing to take
this steepening into account would cause us to underestimate the low-energy extragalactic surface
brightness and thus overestimate the halo surface brightness.  We estimated the size of this
systematic effect by using the double broken power-law model for the extragalactic background from
\citet{smith07a}. Both broken power-laws have $\Gamma_2 = 1.4$ above the break energy of
1.2~\kev. The first component has $\Gamma_1 = 1.54$ below 1.2~\kev, and a normalization of
5.70~\pownorm. The second component has $\Gamma_1 = 1.96$ and a nominal normalization of
4.90~\pownorm. Note that the second normalization was a free parameter in \citepossessive{smith07a}
analysis, and they obtained a value roughly 50\%\ lower. However, because of the aforementioned
degeneracy between the extragalactic background and the soft proton component, we fix the
normalization at is nominal value. Note also that the spectrum of the extragalactic background
depends on the flux level to which sources are removed; it is not clear to what source removal
threshold the \citet{smith07a} model is applicable. Nevertheless, we proceeded by repeating our
analysis for each sightline, using this new extragalactic background model. As with the
normalization of the extragalactic background, we used the differences between the original results
and these new results to estimate the systematic errors for each sightline. The systematic errors on
the halo temperature are typically less than $\pm0.1 \times 10^6~\K$. As expected, this new
extragalactic model generally yields lower halo emission measures, although the difference is
typically less than 0.2~dex.

To calculate the systematic errors on the halo parameters for each sightline, we added the three
systematic errors discussed above in quadrature. We report these combined systematic errors
alongside the statistical errors in the following section.

\subsection{Spectral Fit Results}
\label{subsec:FitResults}

\begin{deluxetable*}{rlccccr@{/}lc}
\tabletypesize{\scriptsize}
\tablewidth{0pt}
\tablecaption{Spectral Fit Results\label{tab:FitResults}}
\tablehead{
& \colhead{Obs. ID}	& \colhead{Foreground R12 rate}		& \colhead{\NH}			& \colhead{Halo $T$}		& \colhead{Halo E.M.}			& \multicolumn{2}{c}{$\chisq/\mathrm{dof}$}     & \colhead{\Stotal} \\
&			& \colhead{(\rosat\ units)}		& \colhead{($10^{20}~\pcmsq$)}	& \colhead{($10^6~\K$)}		& \colhead{($10^{-3}~\emismeas$)}	& \multicolumn{2}{c}{}                          & \colhead{($10^{-12}~\flux~\pdegsq$)} \\
& \colhead{(1)}		& \colhead{(2)}				& \colhead{(3)}			& \colhead{(4)}			& \colhead{(5)}				& \multicolumn{2}{c}{(6)}                       & \colhead{(7)} \\
}
\startdata
 1 & 0304070501 &        743 &      0.790 & $2.31^{+0.08}_{-0.24}\,^{+0.16}_{-0.24}$ & $1.63^{+0.20}_{-0.25}\,^{+0.29}_{-0.64}$ &     405.97 &        362 &       1.66 \\
 2 & 0305290201 &        293 &       5.83 & $2.19^{+0.07}_{-0.08}\,^{+0.21}_{-0.16}$ & $3.96^{+0.27}_{-0.45}\,^{+0.78}_{-0.76}$ &     475.90 &        440 &       3.73 \\
 3 & 0401210601 &        593 &       3.45 & $2.25^{+0.11}_{-0.23}\,^{+0.16}_{-0.21}$ & $1.80^{+0.27}_{-0.34}\,^{+0.45}_{-0.52}$ &     452.03 &        395 &       1.75 \\
 4 & 0404220101 &        606 &       1.72 & $1.95^{+0.20}_{-0.19}\,^{+0.72}_{-0.17}$ & $1.78^{+0.18}_{-0.52}\,^{+0.32}_{-1.35}$ &     469.00 &        422 &       1.40 \\
 5 & 0400560301 &        643 &       1.60 &          $2.10^{+0.05}_{-0.06} \pm 0.16$ & $3.05^{+0.17}_{-0.19}\,^{+0.56}_{-0.64}$ &     693.09 &        594 &       2.71 \\
 6 & 0400570201 &        586 &      0.642 & $2.08^{+0.09}_{-0.13}\,^{+0.17}_{-0.24}$ & $1.79^{+0.23}_{-0.19}\,^{+0.40}_{-0.53}$ &     504.83 &        529 &       1.57 \\
 7 & 0406630201 &        577 &       1.00 & $11.18^{+1.04}_{-0.50}\,^{+0.16}_{-1.41}$ & $1.64^{+0.16}_{-0.49}\,^{+0.29}_{-0.56}$ &     285.43 &        286 &       1.89 \\
 8 & 0303260201 &        555 &      0.640 & $1.83^{+0.02}_{-0.06}\,^{+0.16}_{-0.17}$ &          $4.49^{+0.14}_{-0.33} \pm 1.01$ &     607.54 &        593 &       3.05 \\
 9 & 0306060201 &        687 &       1.26 & $1.46^{+0.03}_{-0.04}\,^{+0.49}_{-0.16}$ & $5.54^{+2.27}_{-0.52}\,^{+0.98}_{-3.35}$ &     587.85 &        594 &       2.13 \\
10 & 0306060301 &        687 &       1.26 & $2.06^{+0.08}_{-0.13}\,^{+0.16}_{-0.21}$ & $3.01^{+0.16}_{-0.49}\,^{+0.54}_{-0.57}$ &     405.23 &        412 &       2.60 \\
11 & 0303720601 &        641 &       1.21 & $2.11^{+0.12}_{-0.09}\,^{+0.32}_{-0.17}$ & $2.16^{+0.26}_{-0.25}\,^{+1.09}_{-0.97}$ &     498.38 &        445 &       1.93 \\
12 & 0303720201 &        641 &       1.21 & $2.08^{+0.08}_{-0.06}\,^{+0.17}_{-0.16}$ & $3.07^{+0.22}_{-0.36}\,^{+0.88}_{-0.53}$ &     494.35 &        494 &       2.68 \\
13 & 0200960101 &        615 &       1.94 &          $2.20^{+0.05}_{-0.08} \pm 0.17$ & $2.12^{+0.11}_{-0.19}\,^{+0.39}_{-0.45}$ &     606.12 &        594 &       2.01 \\
14 & 0301340101 &        575 &       3.31 &          $1.99^{+0.06}_{-0.11} \pm 0.16$ & $3.35^{+0.29}_{-0.42}\,^{+0.74}_{-0.64}$ &     349.15 &        403 &       2.73 \\
15 & 0306370601 &        742 &      0.656 & $1.72^{+0.12}_{-0.10}\,^{+0.16}_{-0.18}$ & $3.55^{+0.35}_{-0.68}\,^{+0.81}_{-0.58}$ &     353.40 &        340 &       2.11 \\
16 & 0402780701 &        471 &       4.54 & $2.03^{+0.37}_{-0.48}\,^{+9.89}_{-0.20}$ & $0.69^{+0.37}_{-0.36}\,^{+0.56}_{-0.45}$ &     400.90 &        451 &       0.58 \\
17 & 0304203401 &        763 &       1.19 & $1.96^{+0.09}_{-0.10}\,^{+0.17}_{-0.16}$ & $4.97^{+0.47}_{-0.54}\,^{+0.88}_{-1.16}$ &     271.14 &        239 &       3.91 \\
18 & 0406610101 &        806 &       1.65 & $3.14^{+0.42}_{-0.23}\,^{+0.61}_{-0.22}$ & $2.45^{+0.23}_{-0.69}\,^{+0.44}_{-0.92}$ &     367.42 &        368 &       3.25 \\
19 & 0400830301 &        527 &       1.74 & $1.91^{+0.10}_{-0.06}\,^{+0.22}_{-0.18}$ & $2.39^{+0.28}_{-0.17}\,^{+0.79}_{-0.78}$ &     630.86 &        594 &       1.80 \\
20 & 0301651701 &        542 &       1.73 &                 $1.83 \pm 0.05 \pm 0.16$ & $6.79^{+0.37}_{-0.46}\,^{+1.20}_{-1.39}$ &     386.43 &        376 &       4.65 \\
21 & 0300630301 &        698 &       2.13 & $1.99^{+0.26}_{-0.36}\,^{+0.17}_{-0.19}$ & $1.03^{+0.43}_{-0.32}\,^{+0.58}_{-0.32}$ &     380.59 &        413 &       0.84 \\
22 & 0312190601 &        427 &       2.29 &          $2.28^{+0.13}_{-0.21} \pm 0.19$ & $1.85^{+0.25}_{-0.39}\,^{+0.73}_{-0.72}$ &     322.29 &        326 &       1.84 \\
23 & 0301330401 &        417 &       2.73 & $2.19^{+0.11}_{-0.16}\,^{+0.41}_{-0.16}$ & $1.63^{+0.25}_{-0.24}\,^{+0.36}_{-0.84}$ &     457.39 &        496 &       1.54 \\
24 & 0312190701 &        573 &       1.75 & $2.04^{+0.07}_{-0.10}\,^{+0.16}_{-0.19}$ & $3.78^{+0.31}_{-0.33}\,^{+1.04}_{-0.85}$ &     375.97 &        363 &       3.21 \\
25 & 0302500101 &        597 &       3.01 & $2.35^{+0.03}_{-0.06}\,^{+0.19}_{-0.16}$ & $5.27^{+0.13}_{-0.36}\,^{+0.94}_{-1.50}$ &     446.89 &        476 &       5.48 \\
26 & 0307001401 &        666 &       2.63 & $3.31^{+0.90}_{-1.09}\,^{+7.62}_{-0.47}$ & $0.37^{+0.19}_{-0.23}\,^{+0.07}_{-0.29}$ &     380.72 &        335 &       0.51 \\

\enddata
\tablecomments{Column~1 contains the \xmm\ observation ID.
Column~2 contains the foreground R12 (1/4~\kev) count-rate in \rosat\ units ($10^{-6}~\rassrate$).
This count-rate was derived from the \citet{snowden00} catalog of SXRB shadows, and was used to fix the normalization of the $1.2 \times 10^6$-\K\ foreground component.
Column~3 contains the \HI\ column density \citep{kalberla05} that was used to attenuate the halo and extragalactic components.
Columns 4 and 5 contain the best-fit halo parameters ($\mbox{E.M.} = \int \Ne^2 dl$). In each case, the first error indicates the
$1\sigma$ statistical error, and the second error indicates the estimated systematic error due to our assumed foreground and extragalactic spectra
(see Section~\ref{subsec:Systematics} for details).
Column 6 contains $\chi^2$ and the number of degrees of freedom.
Column 7 contains the intrinsic 0.4--2.0~\kev\ surface brightness of the halo component.
}
\end{deluxetable*}

The results of our spectral analysis are presented in Table~\ref{tab:FitResults}. In particular,
columns~4 and 5 give the best-fitting halo temperature and emission measure (E.M.) for each
observation, and column~7 gives the intrinsic 0.4--2.0~\kev\ surface brightness of the halo
component. For the temperature and the emission measure, we present both the $1\sigma$ statistical
errors and the estimated systematic errors discussed in the previous section.

\begin{figure*}
\centering
\plottwo{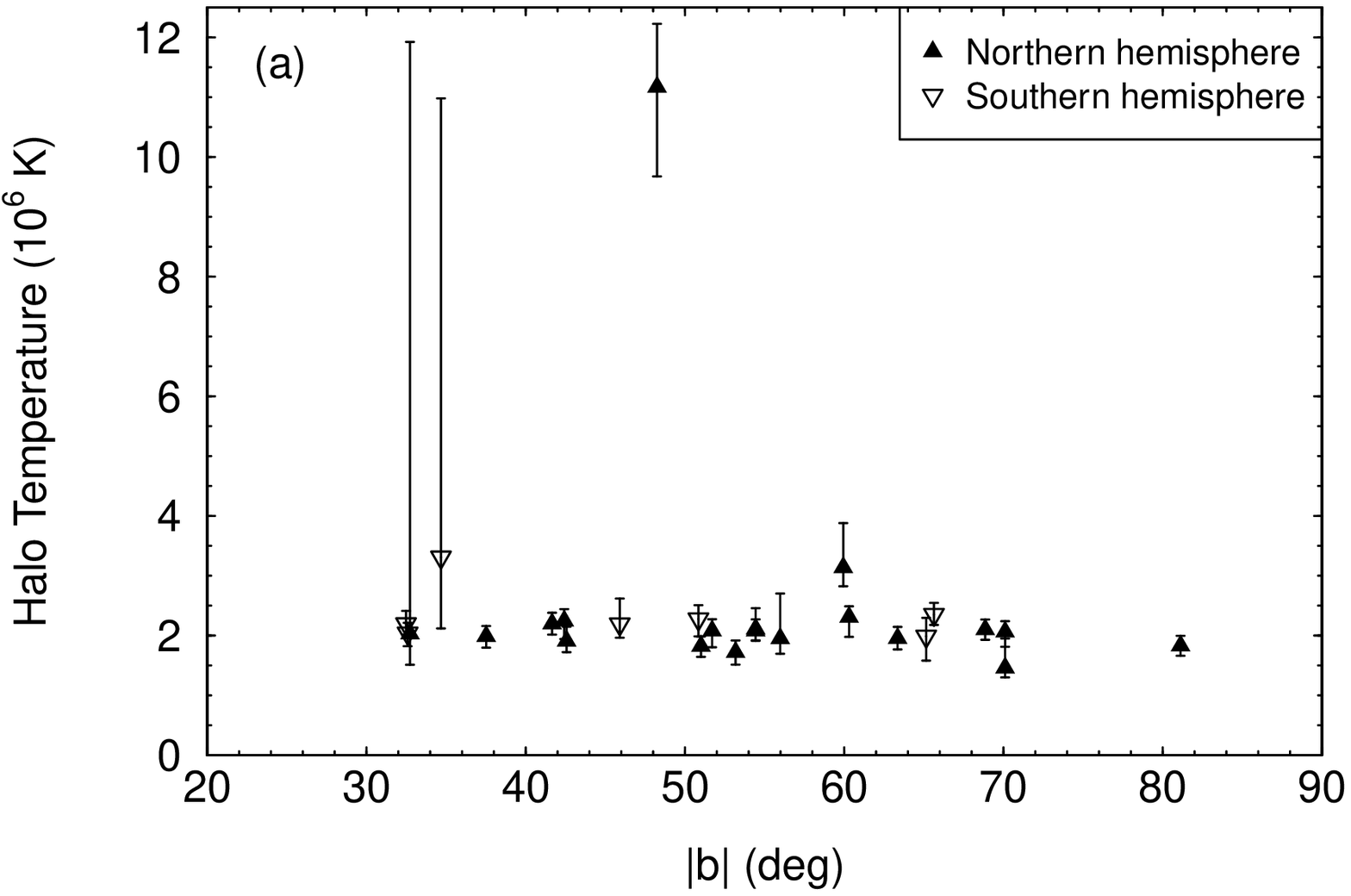}{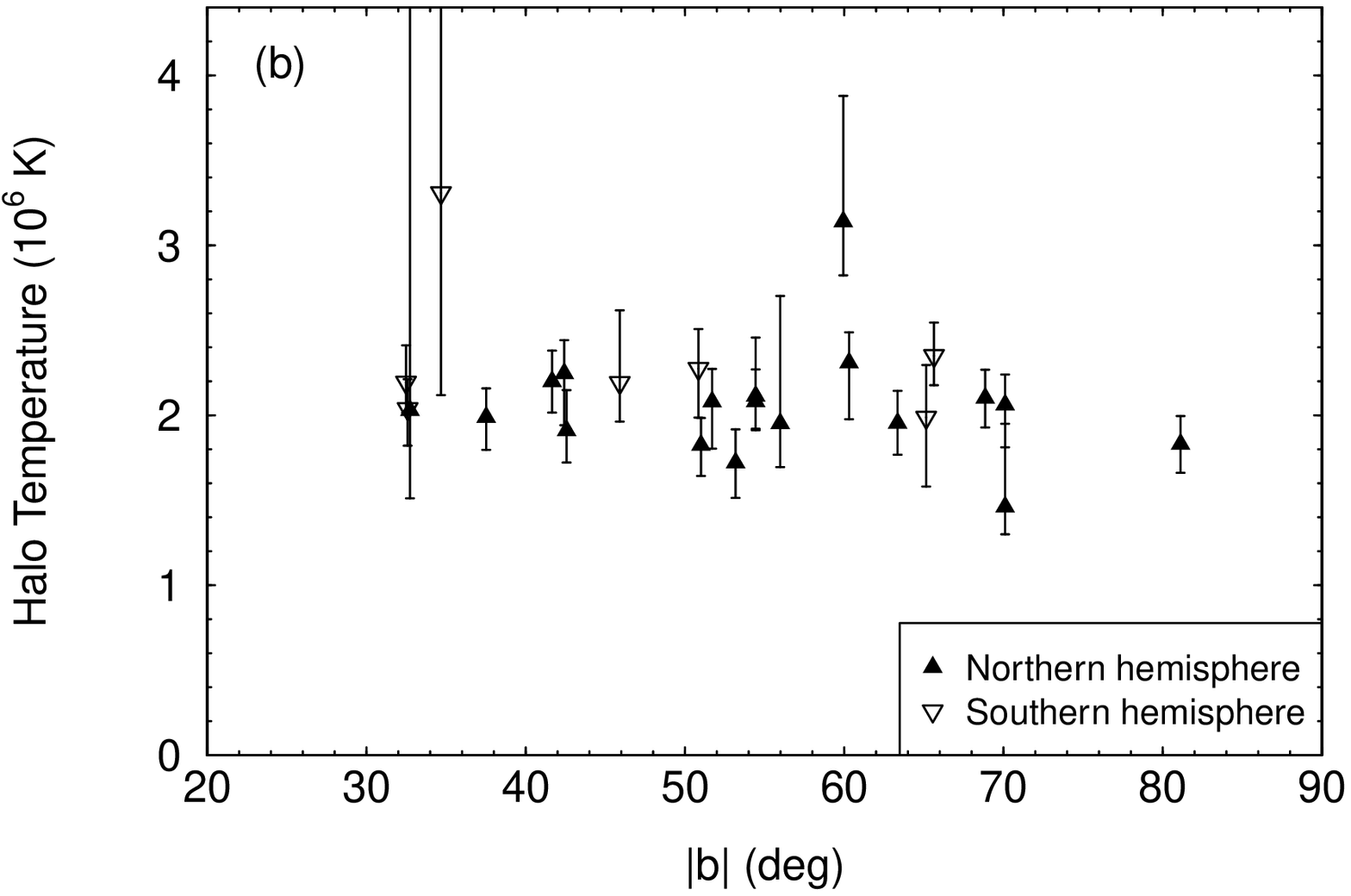}
\plottwo{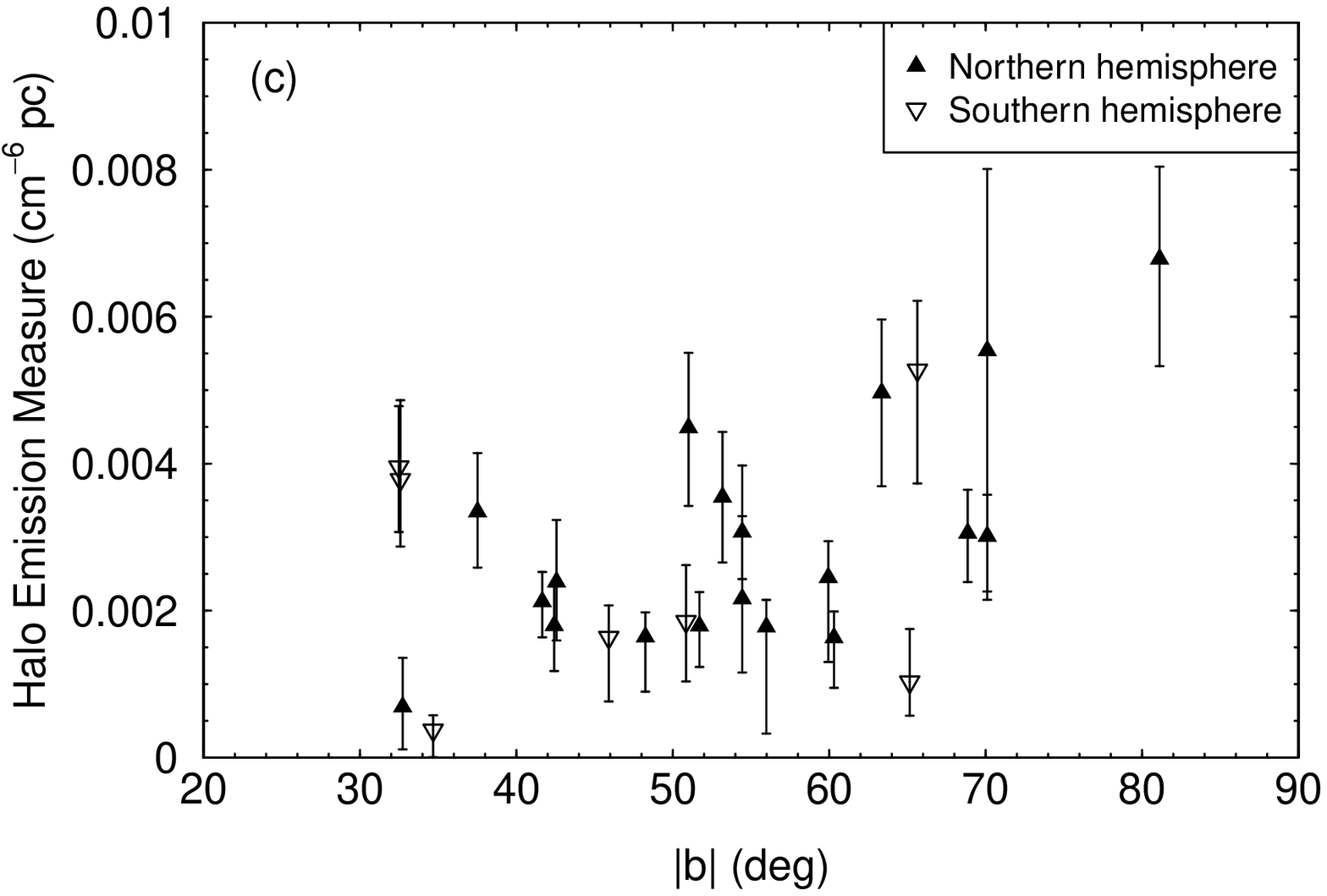}{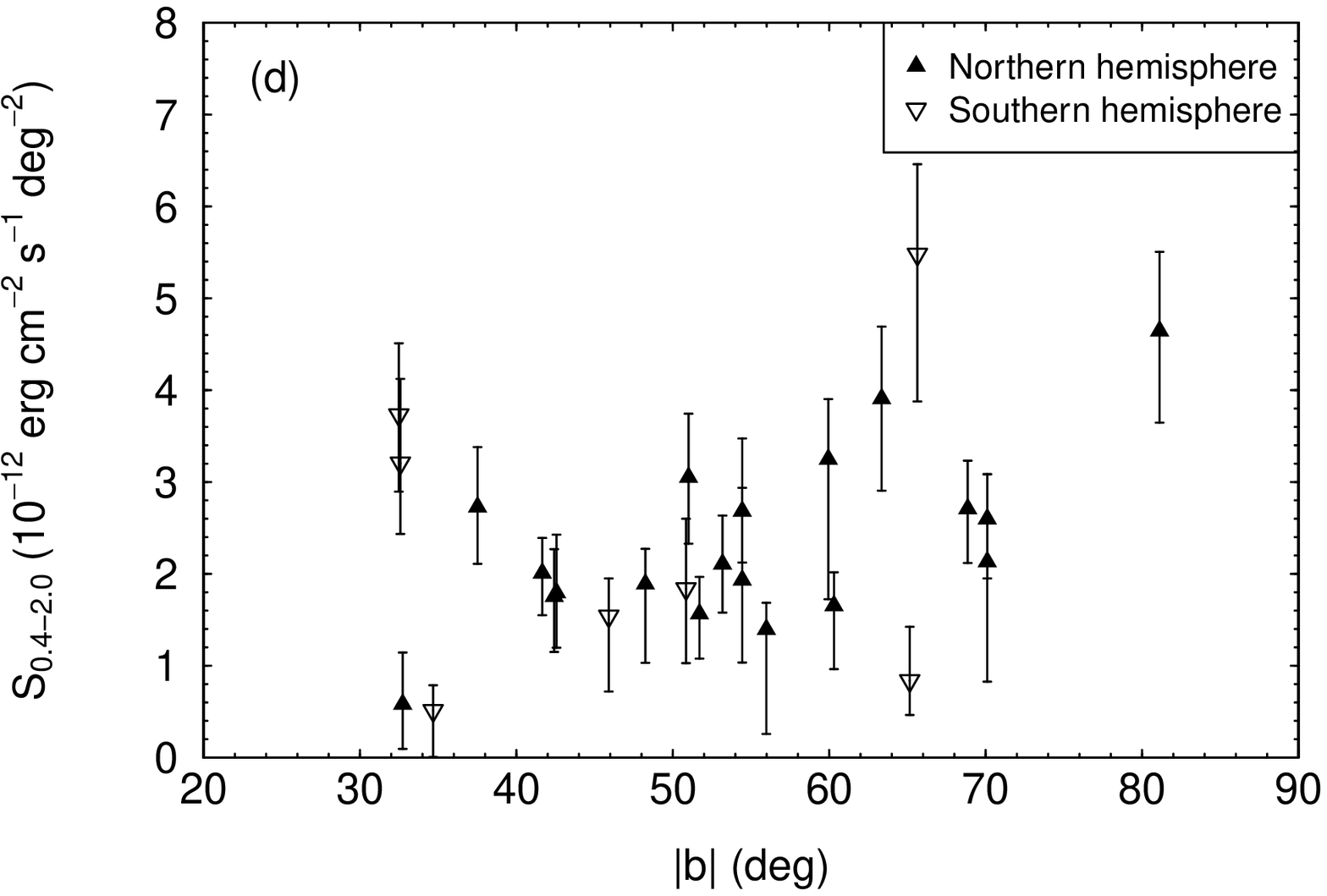}
\caption{Halo (a,b) temperature, (c) emission measure, and (d) intrinsic 0.4--2.0~\kev\ surface
  brightness plotted against Galactic latitude.  Panel~(b) shows the same data as panel~(a), but
  with a narrower $y$-axis range. The error bars show the statistical and systematic errors
  added in quadrature. The errors on the surface brightness are derived from the
  errors on the emission measure.
  \label{fig:Results-vs-abs}}
\end{figure*}

The best-fit halo parameters are plotted against Galactic latitude in Figure~\ref{fig:Results-vs-abs}.
The error bars show the statistical and systematic errors added in quadrature. The
observations with large temperature error bars at $|b| \approx 33\degr$ and 35\degr\ are 0402780701
and 0307001401 (numbers~16 and 26, respectively, in Tables~\ref{tab:ObservationDetails} and
\ref{tab:FitResults}). Note that the halo is faint in these directions (indeed, these observations
yield the two lowest 0.4--2.0~\kev\ surface brightnesses), and that the exposure times are not
unusually long. It is therefore unsurprising that the halo temperatures are poorly constrained for
these observation. The other temperatures are generally well constrained; they are typically
$\sim$$(\mbox{1.8--2.4}) \times 10^6$~K, and are fairly constant across the sky. The halo emission
measures mostly lie in the range $\sim$0.0005--0.006~\emismeas, although there is significant
variation within that range. Correspondingly, there is also significant variation in the intrinsic
X-ray surface brightness, within the range $\sim$$(\mbox{0.5--5}) \times 10^{-12}~\flux\ \pdegsq$.

\begin{figure*}
\centering
\plottwo{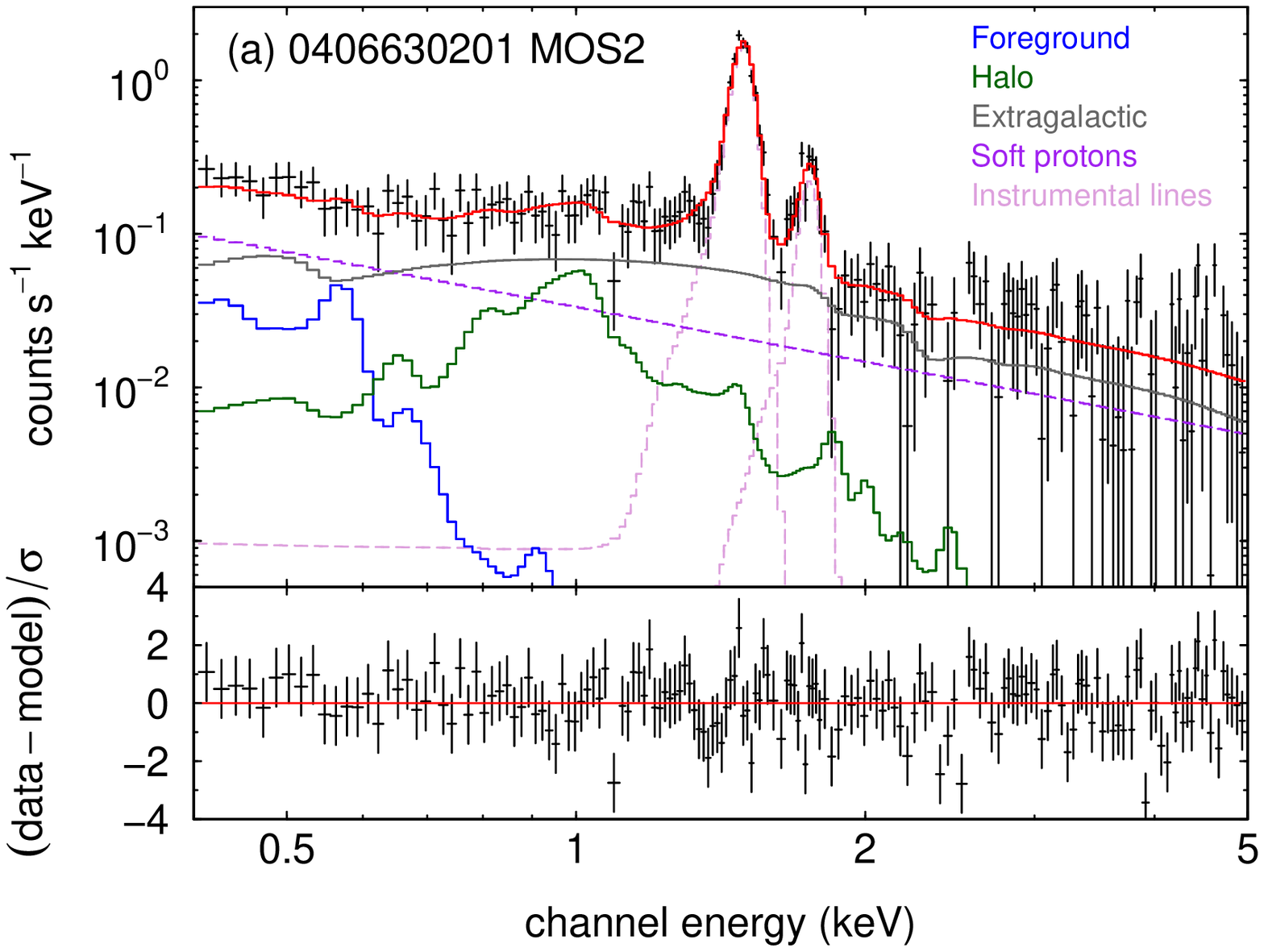}{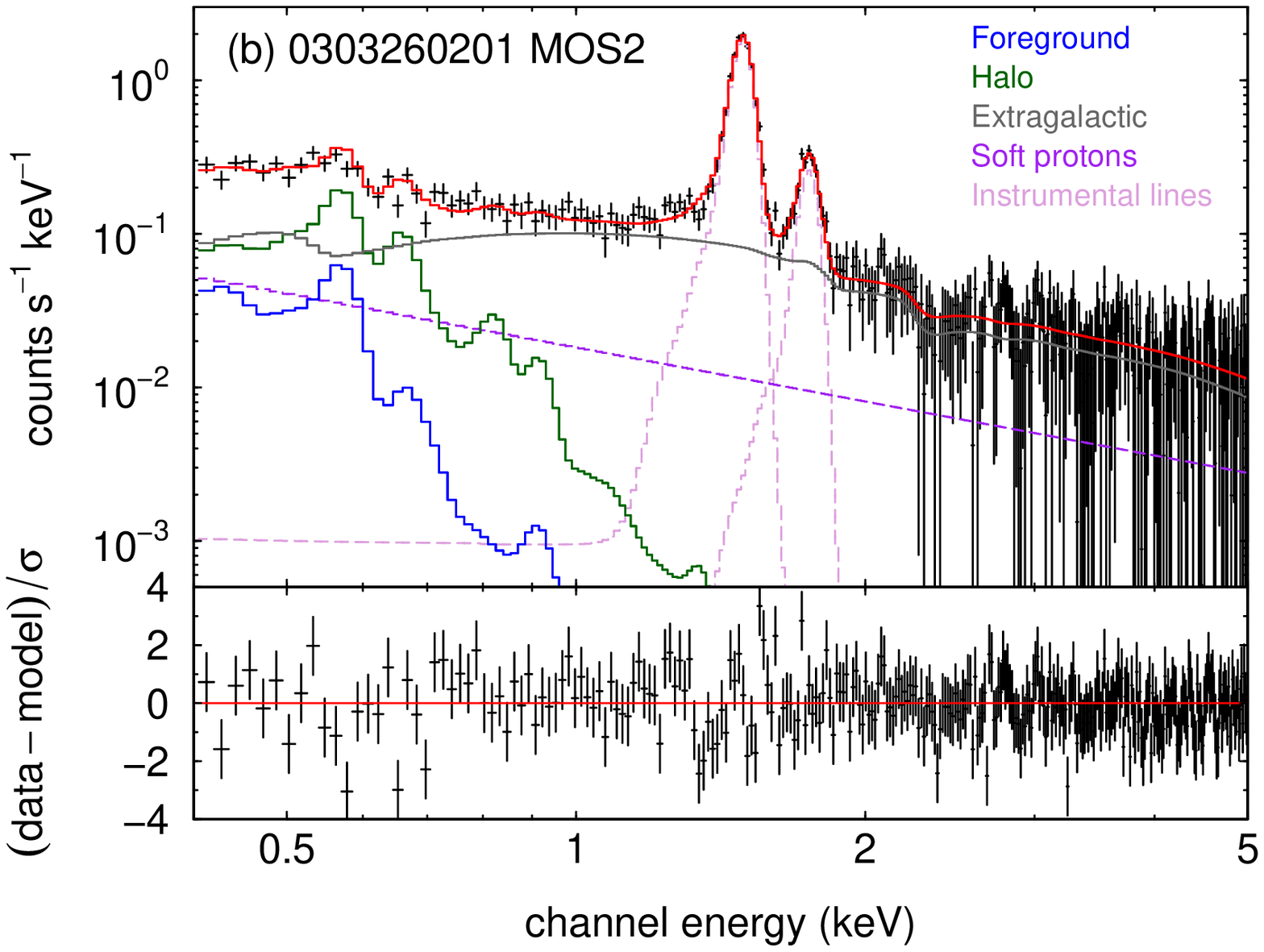}
\caption{MOS2 spectra from (a) the observation with an exceptionally high halo temperature of $11.2
  \times 10^6$~\K, obs.~0406630201, and (b) a more normal observation, obs.~0303260201, showing the
  best-fitting model (in red) and the individual model components. Components of the particle
  background are plotted with dashed lines.
  \label{fig:Spectra}}
\end{figure*}

One observation, 0406630201 (number~7 in Table~\ref{tab:FitResults}), gives a significantly higher
halo temperature than the other observations ($11.2 \times 10^6$ versus $\sim$$2 \times
10^6$~\K). The MOS2 spectrum and best-fit model for this observation are shown in
Figure~\ref{fig:Spectra}. For comparison, Figure~\ref{fig:Spectra} also shows the spectrum from a
nearby observation (0303260201; number~8 in Table~\ref{tab:FitResults}) that yields a more typical
halo temperature ($1.8 \times 10^6$~\K). The halo component for obs.~0406630201 is faint, and the
temperature may be less well constrained than the formally calculated error bar implies. We tried
repeating our analysis of this observation with the halo temperature fixed at $2 \times 10^6$~K
(i.e., similar to the temperatures found from most of the other observations). The best-fit emission
measure for this new halo component was essentially zero ($3\sigma$ upper limit:
0.0022~\emismeas). This new model yielded an acceptable fit: $\chi^2 = 309.59$ for 287 degrees of
freedom. Therefore, although the formal best-fitting temperature is $11.2 \times 10^6$~K, this
spectrum is also consistent with a $\sim$$2 \times 10^6$~K halo with a small emission measure
($\la$0.002~\emismeas). The upper limit on the emission measure is not unusually small compared with
some of the other observations.

Although there is this one anomalous temperature among our 26 \xmm\ observations, it should not affect
our subsequent analysis.  In Section~\ref{sec:ComparisonWithModels}, we will compare our
observations of the halo with the predictions of various physical models. Rather than looking at
individual sightlines, we look at the whole population of observational results, and compare
histograms of observational properties with corresponding histograms derived from the
models. Therefore, as long as the majority of our observations yield reasonably accurate halo
properties, a single outlying anomalous result should not significantly affect the comparison
of the observations with the models.

\section{COMPARING THE OBSERVED HALO X-RAY SPECTRA WITH HYDRODYNAMICAL MODELS}
\label{sec:ComparisonWithModels}

In this section, we compare the X-ray spectral properties of the halo inferred from our
\xmm\ observations with those predicted by various hydrodynamical models. In particular, in
Section~\ref{subsec:DiskGalaxyFormation} we examine a disk galaxy formation model, in which
extragalactic material is heated as it falls into the Galaxy's potential well
\citep{toft02,rasmussen09}. In Section~\ref{subsec:InSituSupernovae} we examine a model in which the
hot halo gas is heated \textit{in situ} by isolated extraplanar SNe \citep{shelton06}.  In
Section~\ref{subsec:SNDrivenISM} we examine a model in which the ISM is heated and stirred by
multiple SNe \citep{joung06}. Unlike the previous model, the SNRs are not assumed to evolve in
isolation, and the model includes the movement of hot gas from the disk into the halo.

\subsection{Disk Galaxy Formation Model}
\label{subsec:DiskGalaxyFormation}

Cosmological smoothed particle hydrodynamics (SPH) simulations of disk galaxy formation predict that
such galaxies should be surrounded by extended hot halos ($r \sim 10$s of \kpc, $T \sim \mbox{few}
\times 10^6$~\K; \citealt{toft02,rasmussen09}). These halos contain a significant fraction of the
galactic baryonic mass \citep{sommerlarsen06}. In this subsection we compare the predictions of such models
with our \xmm\ observations.

\begin{figure}
\centering
\plotone{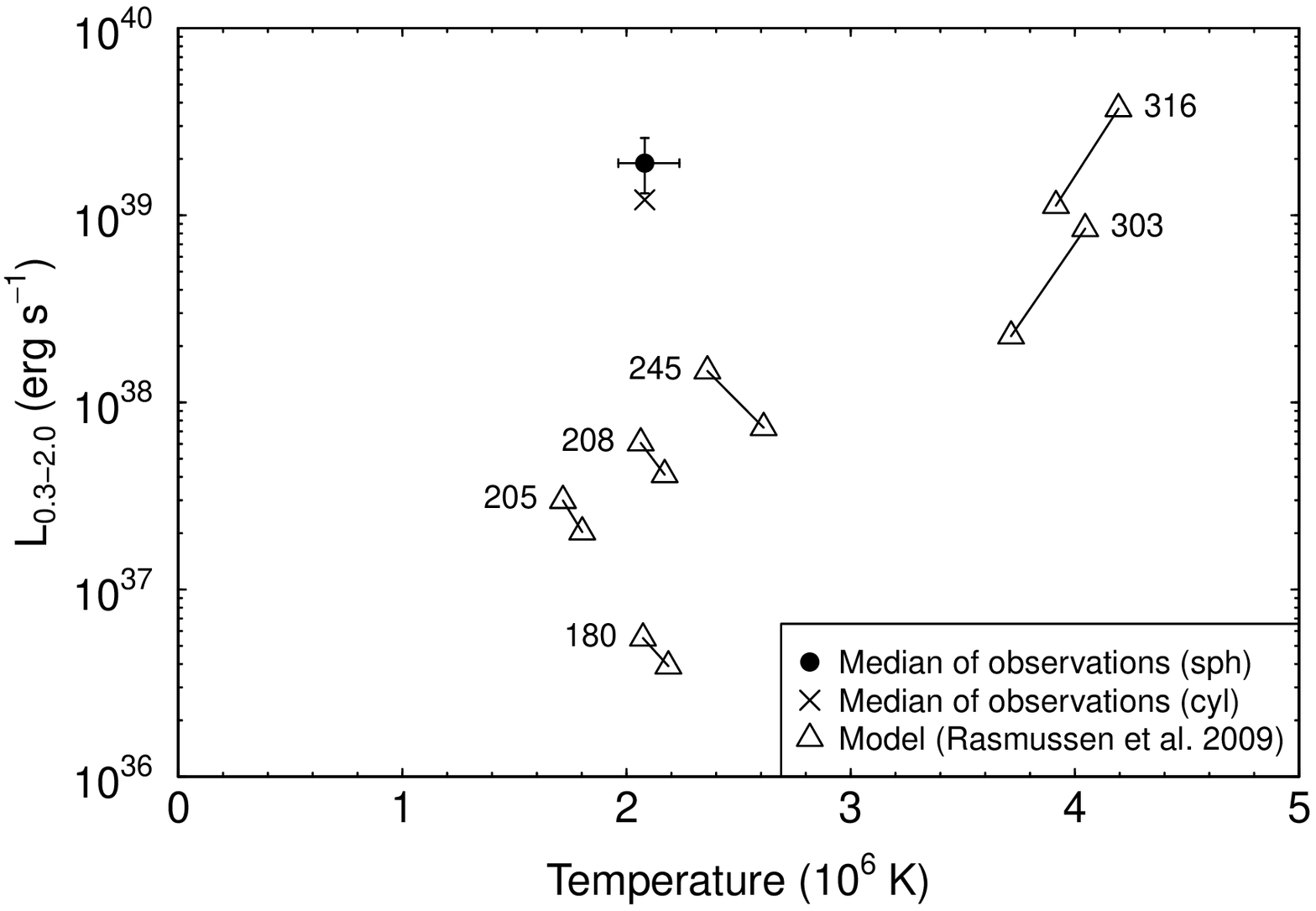}
\caption{Halo temperatures and 0.3--2.0~\kev\ luminosities derived from our \xmm\ observations and
  from disk galaxy formation simulations \citep{rasmussen09}. The triangles show the
  emission-weighted mean temperatures and halo luminosities predicted by the model. These datapoints
  are labeled with the model galaxies' circular velocities, \vc, in \kmps. For each model galaxy,
  the lower datapoint shows the values extracted from within a spherical region of radius 40~\kpc,
  and the upper datapoint shows the values extracted from within a $(100~\kpc)^3$ box. In both cases, a
  cylindrical region around the galactic disk was excluded. The solid circle and the diagonal cross show the
  median observed values.  The solid circle denotes the median halo luminosity derived from the
  observations assuming a spherical emission geometry (Equation~(\ref{eq:HaloLxSph})); the errorbars
  indicate the lower and upper quartiles. The diagonal cross
  denotes the median halo luminosity derived assuming a cylindrical emission geometry
  (Equation~(\ref{eq:HaloLxCyl})).
  \label{fig:DiskGalaxy+Obs}}
\end{figure}

J. Rasmussen (2009, private communication) has kindly provided us with 0.3--2.0~\kev\ luminosities
and emission-weighted mean temperatures derived from the SPH galaxy formation simulations described
in \citet{rasmussen09}. \citet{rasmussen09} point out that the X-ray emission from the hot gas
particles can be artificially boosted by nearby cold, dense gas particles (within an SPH smoothing
length, \hSPH). As $\hSPH \approx 1.5~\kpc$ in these simulations, the emission from the disk can be
particularly adversely affected. Therefore, a cylindrical region around each galactic disk within
$|z| = 2~\kpc$ and $r = 15~\kpc$ was excluded from the calculation of the X-ray properties.  For
each galaxy, two sets of values were extracted: one extracted from within a spherical region of
radius 40~\kpc\ (these are the values shown in Figures~5 and 6 in \citealp{rasmussen09}), and one
extracted from within a $(100~\kpc)^3$ box. The X-ray luminosities and temperatures for each model galaxy
are shown by the triangles in Figure~\ref{fig:DiskGalaxy+Obs}.

In order to compare the model predictions with our observations, we must first derive a luminosity
for the Galactic halo from our observations. To do this, we must assume some geometry for the halo
emission. In the following, we consider both spherical and cylindrical geometries. Of the two, a
spherical geometry is probably the more appropriate for comparison with the extended hot halo
predicted by the model, especially as a cylindrical region around each model galactic disk was
excluded before the X-ray properties were extracted.

For the spherical halo geometry, we assume that the halo emission comes from a uniform sphere of
radius \Rsph. If the intrinsic surface brightness, \SX, along some sightline is typical for the whole
galaxy, the luminosity per unit volume of the halo is $4 \pi \SX / \lambda$, where $\lambda$ is the
path length through the spherical halo. This path length is a function of the viewing direction:
\begin{equation}
  \lambda = \Rsol \cos l \cos b + \sqrt{\Rsph^2 - (1 - \cos^2 l \cos^2 b) \Rsol^2},
\end{equation}
where $\Rsol = 8.5~\kpc$ is the radius of the solar circle. The X-ray luminosity is then given by
\begin{equation}
  \LX = \frac{16 \pi^2}{3} \frac{\Rsph^3 \SX}{\lambda}.
  \label{eq:HaloLxSph}
\end{equation}

For the cylindrical halo geometry, if the intrinsic surface brightness along some sightline is
typical for the whole galaxy, the luminosity per unit area of the halo
integrated over the vertical direction is $2 \times 4 \pi \sin(|b|)
\SX$. The initial factor of 2 takes into account the halo above and below the disk. If we then
assume that the halo emission originates within a cylindrical region of radius \Rcyl, the total
X-ray luminosity of the Galactic halo is
\begin{equation}
  \LX = 8 \pi^2 \sin(|b|) \Rcyl^2 \SX.
  \label{eq:HaloLxCyl}
\end{equation}

We have calculated intrinsic 0.3--2.0~\kev\ surface brightnesses for the halo from our best-fitting
spectral models in Table~\ref{tab:FitResults}. We then converted each surface brightness to a
0.3--2.0~\kev\ luminosity, using Equations~(\ref{eq:HaloLxSph}) and (\ref{eq:HaloLxCyl}). In both cases
we assumed emission radii (\Rsph\ or \Rcyl) of 15~\kpc. There is a large amount of scatter in the
derived luminosities, spanning about an order of magnitude. The medians of these values are $1.9
\times 10^{39}~\ergps$ (spherical geometry) and $1.2 \times 10^{39}~\ergps$ (cylindrical geometry);
these are our best estimates of the 0.3--2.0~\kev\ luminosity of the Milky Way halo. However, it should
be noted that our sightlines are all in directions away from the Galactic Center, and so if the halo
emissivity increases toward the Galactic Center, the above values will underestimate the Galactic
luminosity. In addition, the luminosity inferred assuming a spherical geometry will be an underestimate
if the halo is more extended than our assumed $\Rsph = 15~\kpc$ (note from Figure~\ref{fig:DiskGalaxy+Obs}
that the model predicts that a significant fraction of the halo emission originates from $r > 40~\kpc$).
For comparison, analysis of \rosat\ All-Sky Survey data has yielded halo X-ray luminosities of $7
\times 10^{39}~\ergps$ in the 0.1--2.0~\kev\ band (assuming that the quoted value applies to the
whole \rosat\ band; \citealp{pietz98}) and $3 \times 10^{39}~\ergps$ in the 0.5--2.0~\kev\ band
\citep{wang98}. Assuming a halo temperature of $2 \times 10^6~\K$, these values correspond to 0.3--2.0~\kev\
luminosities of $4 \times 10^{39}$ and $5 \times 10^{39}~\ergps$, respectively.

Figure~\ref{fig:DiskGalaxy+Obs} compares the median observed halo temperature and the
above-mentioned median halo luminosities with the predictions of the disk galaxy formation
simulations. For model galaxies similar in size to the Milky Way ($\vc \approx 220~\kmps$), the
predicted temperatures are in good agreement with the observations. However, the \citet{rasmussen09} model
underpredicts the observed halo luminosity by at least an order of magnitude.  We will discuss these results
in Section~\ref{subsubsec:DiskGalaxyFormationDiscussion}.

\subsection{Extraplanar Supernova Explosions}
\label{subsec:InSituSupernovae}

Here, we consider a model in which the hot halo gas is heated locally by SNe above the Galactic
disk. In this scenario, the observed hot gas is within isolated SNRs at a variety of heights above
the disk and of a variety of ages. Superbubbles blown by clustered SNe and hot gas that has risen
from the disk are excluded in this model. This model was developed by \citet{shelton06}, who found
that it could explain a significant fraction of the high-latitude 1/4~\kev\ halo emission (excluding
anomalously bright features such as the North Polar Spur). Here, we compare the model to the
higher-energy emission observed with \xmm.

\subsubsection{Model Description}

The model spectra discussed here were generated from one-dimensional Lagrangian hydrodynamical
simulations of SNRs evolving at a variety of heights above the Galactic disk, and hence in a variety of
ambient densities.  The simulations are described fully in \citet{shelton98,shelton99,shelton06},
and include radiative cooling, thermal conduction, and an effective ambient magnetic field, \Beff,
which exerts a non-thermal pressure, in addition to the ambient gas pressure. The
ionization evolution in the shocked gas is modeled self-consistently.

\begin{figure}
\centering
\plotone{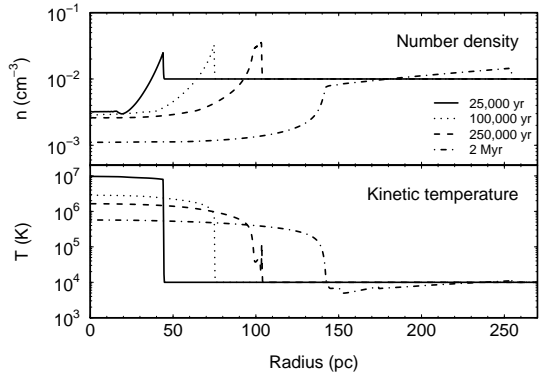}
\caption{Density and temperature profiles for an SNR model with $E_0 = 0.5 \times 10^{51}~\erg$,
  $\Beff = 2.5~\microgauss$ evolving in an ambient medium with $n_0 = 0.01~\pcc$. Note that the cool
  shell has formed by 250,000~\yr.
  \label{fig:SNRprofiles}}
\end{figure}

We will first concentrate on the models from \citet{shelton06} with SN explosion energy $E_0 = 0.5 \times
10^{51}~\erg$ and $\Beff = 2.5~\microgauss$ (corresponding to a non-thermal pressure $\Pnt =
1800~\presalt$), and then discuss varying these parameters. We consider the models
evolving in ambient densities $n_0 = 0.2$, 0.1, 0.05, 0.02, 0.01, and 0.005~\pcc, corresponding to
heights $z = 190$, 310, 480, 850, 1300, and 1800~\pc, using the interstellar density model from
\citet{ferriere98a}. Note that we are ignoring the model from \citet{shelton06} at $z = 76~\pc$,
as we do not consider this to be in the halo.  Density and temperature profiles from the model
with $n_0 = 0.01~\pcc$ are shown in Figure~\ref{fig:SNRprofiles}. The SN explosion blows a hot
rarefied bubble in the ambient medium; this hot bubble is the source of the X-rays. The model
shown in Figure~\ref{fig:SNRprofiles} corresponds to model~A in \citet{shelton99}; see that
paper for more details.

For each epoch of each SNR model, we calculated spectra for a range of impact parameters through the
model remnant, using the \citet{raymond77} spectral code (updated by J.~C. Raymond \& B.~W. Smith,
1993, private communication with R.~J. Edgar). The spacing between the impact parameters, $\Delta
b$, was adjusted according to the size of the remnant. The spectral calculation takes into account
the possible non-equilibrium ionization calculated during the hydrodynamical simulation.

We wish to compare the distributions of observed temperatures and emission measures with those predicted by
the model. In order to do this, we need to calculate the probability of a sightline passing through
a remnant at a given height, of a given age, and at a given impact parameter. For example, a
sightline is more likely to pass through a remnant at a larger impact parameter than a small impact
parameter, and more likely to pass through a remnant closer to the disk, where the SN rate is
larger. If $R(z_1,z_2)$ is the rate per unit area of SNe at heights between $z_1$ and $z_2$, then
the probability $P(z_1,z_2,t,\Delta t,b,\Delta b)$ of intercepting a remnant between these heights with an age between
$t$ and $t + \Delta t$ at an impact parameter between $b$ and $b + \Delta b$ is
\begin{equation}
	P (z_1,z_2,t,\Delta t,b,\Delta b) = 2 \pi b R(z_1,z_2) \Delta b \Delta t.
\label{eq:Probability}
\end{equation}
In order to calculate the above probabilities, we followed \citet{shelton06} and divided the halo into
6 plane-parallel slabs. We assumed that all the SNRs within a given slab experience a single ambient
density, and so can be represented by one of our 6 SNR models. We calculated $R(z_1,z_2)$ for each
slab by integrating the volumetric SN rate at the solar circle, $r(z)$, from \citet{ferriere98a}:
\begin{eqnarray}
        r(z) &=& r_\mathrm{Ia}(z) + r_\mathrm{II}(z)                                            \nonumber \\
             &=& ( 4.0 \e^{-|z|/325~\mathrm{pc}} + 14 \e^{-|z|/266~\mathrm{pc}} ) ~\kpc^{-3}~\Myr^{-1},~~~~
\label{eq:SNrate1}
\end{eqnarray}
where $r_\mathrm{Ia}(z)$ and $r_\mathrm{II}(z)$ are the rates of type Ia SNe and isolated
core-collapse SNe, respectively. The slab boundaries were placed at the midpoints between the
nominal heights of the model remnants, except for the lower boundary of the lowest slab, which was
placed at 130~\pc\ (the scale height of the Galactic \HI\ layer), and the upper boundary of the
highest slab, which was placed at infinity. The boundaries, ambient densities, and SN rates for the
6 slabs are shown in Table~\ref{tab:Slab}.  We calculated $\Delta t$ from the timestamps of the
output files from the hydrodynamical simulations.  As noted above, $\Delta b$ is the spacing between
the impact parameters for which we calculated spectra from a given remnant.

\begin{deluxetable*}{ccccc}
\tablewidth{0pt}
\tablecaption{Densities and Supernova Rates in the Halo as a Function of Height\label{tab:Slab}}
\tablehead{
\colhead{Slab}  & \colhead{Height range}        & \colhead{Ambient density, $n_0$}      & \colhead{Nominal SNR height}  & \colhead{SN rate, $R$} \\
                & \colhead{(pc)}                & \colhead{(\pcc)}                      & \colhead{(pc)}                & \colhead{($\kpc^{-2}~\Myr^{-1}$)} \\
}
\startdata
1               & 130--250                      & 0.2                                   & 190                           & 1.1   \\
2               & 250--395                      & 0.1                                   & 310                           & 0.83  \\
3               & 395--665                      & 0.05                                  & 480                           & 0.76  \\
4               & 665--1075                     & 0.02                                  & 850                           & 0.36  \\
5               & 1075--1550                    & 0.01                                  & 1300                          & 0.091 \\
6               & 1550--$\infty$                & 0.005                                 & 1800                          & 0.022 \\
\enddata
\end{deluxetable*}

We used a Monte Carlo method to construct model spectra for 2000 sightlines, using the
above-calculated probabilities. In our Monte Carlo simulation, 64\%\ of the model sightlines
intercepted no remnants, and 9\%\ of the sightlines intercepted more than 1 remnant. For model
sightlines intercepting more than 1 remnant, we summed the spectra of the individual remnants.

\subsubsection{Characterizing the SNR Spectra with $1T$ Models}
\label{subsubsec:CharacterizingSpectra}

In our analysis of the \xmm\ observations, we modeled the halo X-ray emission with a
single-temperature CIE plasma model, whereas the true halo emission is likely from plasma at a range
of temperatures and in a range of ionization states. Similarly, the emission predicted by this
extraplanar SNR model is from plasma at a range of temperatures and in a range of ionization
states. Therefore, in order to compare the predictions of this model with our observational results,
we first characterize the model SNR spectra calculated above with $1T$ CIE plasma models, by simulating
\xmm\ observations of the SXRB.

Our procedure for characterizing the spectrum for each model sightline is as follows. We first
multiplied the spectrum by a renormalization factor, $k_\mathrm{rn}$, in order to give a
0.4--2.0~\kev\ surface brightness of $2.06 \times 10^{-12}~\flux\ \pdegsq$. This value is the median
intrinsic halo surface brightness inferred from the best-fit models in
Table~\ref{tab:FitResults}. We then subjected the model to an absorbing column $\NH = 1.7 \times
10^{20}~\pcmsq$; again, this is the median value used in our \xmm\ analysis. To the absorbed SNR
spectrum, we added components representing the foreground emission, the extragalactic background,
the soft protons, and the instrumental lines. The parameters for the foreground emission and the
extragalactic background were taken from our observational analysis (see
Section~\ref{subsec:ModelDescription}). The normalization of the foreground component was chosen to
give an R12 count-rate of 600~\rassrate\ (the median value from Table~\ref{tab:FitResults}). The
parameters for the soft proton model and the instrumental lines were taken from the best-fit model
for obs.~0301330401; this observation was chosen because it has close to the median level of soft
proton contamination, as judged by the ratio of the observed 2--5~\kev\ flux to that expected from
the extragalactic background ($F_\mathrm{total}^{2-5} / F_\mathrm{exgal}^{2-5}$; see Paper~I).

We simulated an observation of the model spectrum by folding the above-described multicomponent
model through the \xmm\ response, assuming a typical field of view of 480~arcmin$^2$, and adding
Poissonian noise to the spectrum, assuming a typical observing time of 15~\ks. Our simulations also
took into account the QPB spectrum. For each model sightline, we simulated a MOS1 and a MOS2
spectrum. We binned the resulting spectra such that there were at least 25 counts per bin. Note that
our assumed field of view is not as large as the full \xmm\ field of view, as we removed bright
sources from the fields of our observations, and for some observations not all chips were usable.

\begin{figure}
\centering
\plotone{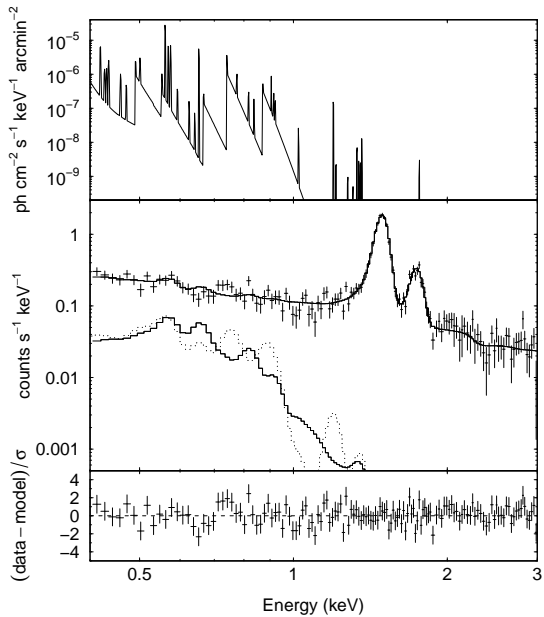}
\caption{Simulated \xmm\ MOS1 spectrum of the SXRB, created using the extraplanar SNR model for the
  halo emission.  The top panel shows the input SNR spectrum. The middle panel shows the simulated
  SXRB spectrum (crosses). The upper solid line shows the best-fit SXRB model, while the lower solid
  line shows the $1T$ halo component from that model (the other model components have been omitted
  for clarity; cf. Figure~\ref{fig:Spectra}). The dotted line shows the input SNR spectrum folded
  through the \xmm\ response. The bottom panel shows the residuals.
  \label{fig:FakeSNRSpectrum}}
\end{figure}

\begin{figure}
\centering
\plotone{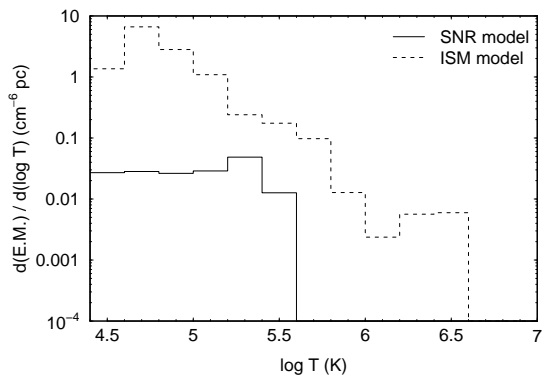}
\caption{
  Differential emission measures for the input SNR model plotted in Figure~\ref{fig:FakeSNRSpectrum} (solid line)
  and the input SN-driven ISM model plotted in Figure~\ref{fig:FakeISMSpectrum} (dashed line).
  \label{fig:DEM}}
\end{figure}

We fitted the resulting simulated spectra with the same multicomponent model that we used in our
observational analysis. In particular, the halo emission was modeled with an absorbed
\citeauthor{raymond77} model, with \NH\ fixed at $1.7 \times 10^{20}~\pcmsq$. For each model
sightline we fitted to the simulated MOS1 and MOS2 spectra simultaneously.
Figure~\ref{fig:FakeSNRSpectrum} shows a simulated \xmm\ spectrum from one of our model sightlines,
along with the best-fitting model. Also shown is the $1T$ CIE halo component of the best-fitting
model, as well as the input SNR model. The input SNR spectrum is very different from that of a CIE
plasma.  The SNR plasma is relatively cool (as shown by the differential emission measure, plotted
as the solid line in Figure~\ref{fig:DEM}) and overionized, and the spectrum exhibits strong
recombination edges. If the SNR plasma were in CIE, instead of recombining, its emission would be 4 orders of
magnitude fainter in the \xmm\ band. However, when we use a $1T$ CIE plasma to model the halo
component in our simulated spectra, the fits are generally good. This is because much of the
spectral detail in the input SNR spectrum is lost when it is combined with the other components of
the SXRB and folded through the \xmm\ response. There is some excess emission at $\sim$0.75~\kev\ in
the simulated spectrum in Figure~\ref{fig:FakeSNRSpectrum}, possibly due to $\mathrm{O}^{+7}
\rightarrow \mathrm{O}^{+6}$ recombinations (the recombination edge is at 0.74~\kev), but it would
not be easy to unambiguously identify such a feature in an observed spectrum as recombination
emission.

We used the resulting fit parameters to compare with the observations. The temperatures were taken
directly from the fits, while the best-fit emission measures were first divided by the relevant
values of $k_\mathrm{rn}$ before comparing with the observations. In what follows, we refer to these
values as ``X-ray temperatures'' and ``X-ray emission measures'', to emphasize that they are derived
from simulated X-ray observations, rather than being derived directly from the hydrodynamical data.

\subsubsection{Comparing the Extraplanar SNR Model with Observations}
\label{subsubsec:SNRModel+Obs}

Our model X-ray emission measures were derived for sightlines looking straight upward from the disk, i.e., toward
$|b| = 90\degr$. Our observations, however, are toward a range of Galactic latitudes, and so will
sample different amounts of halo material. We therefore multiply our observed emission measures by $\sin |b|$
before comparing them with the model predictions. This transformation assumes that the halo is, in a
statistical sense at least, uniform in directions parallel to the disk.

\begin{figure}
\centering
\plotone{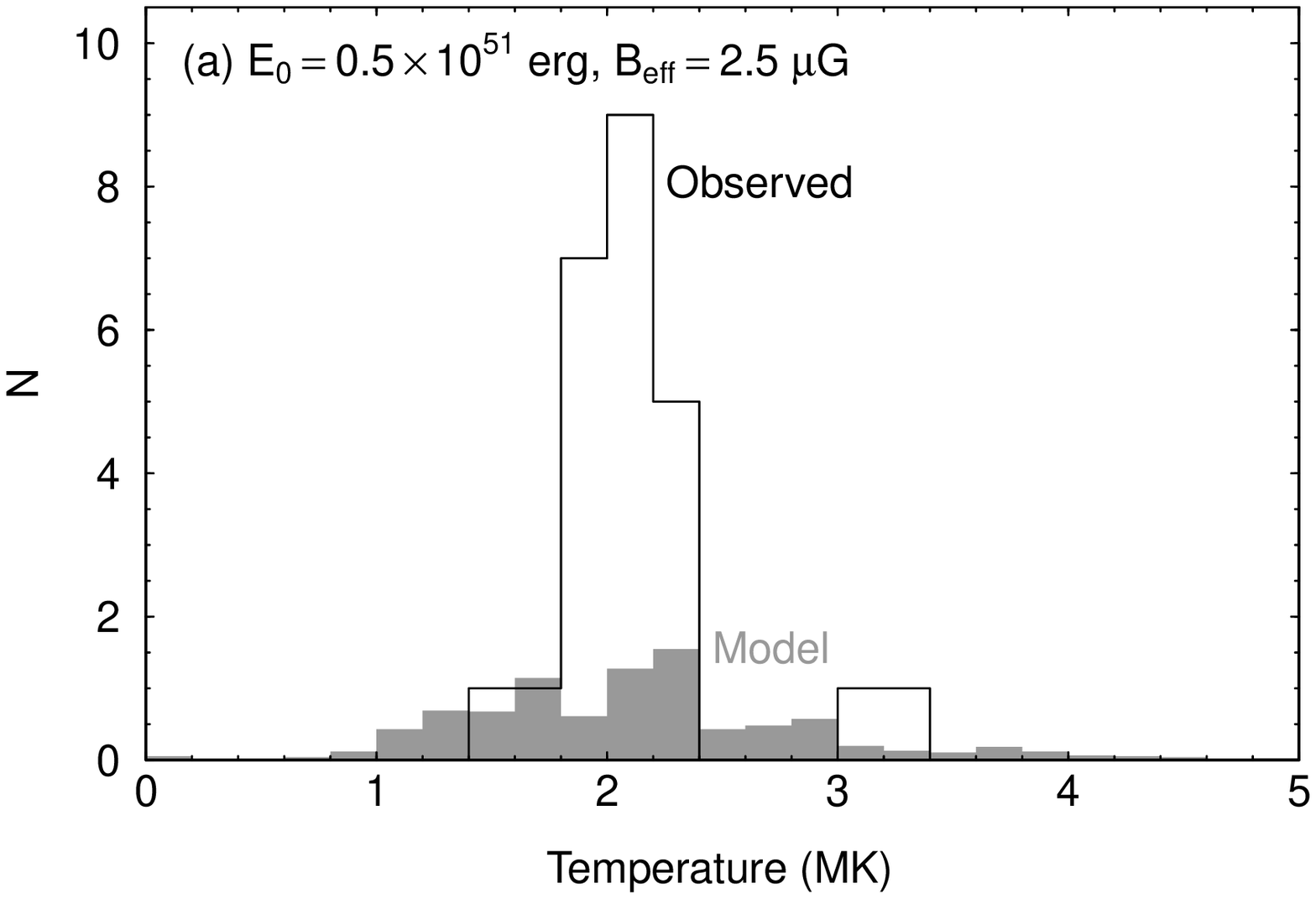} \\
\plotone{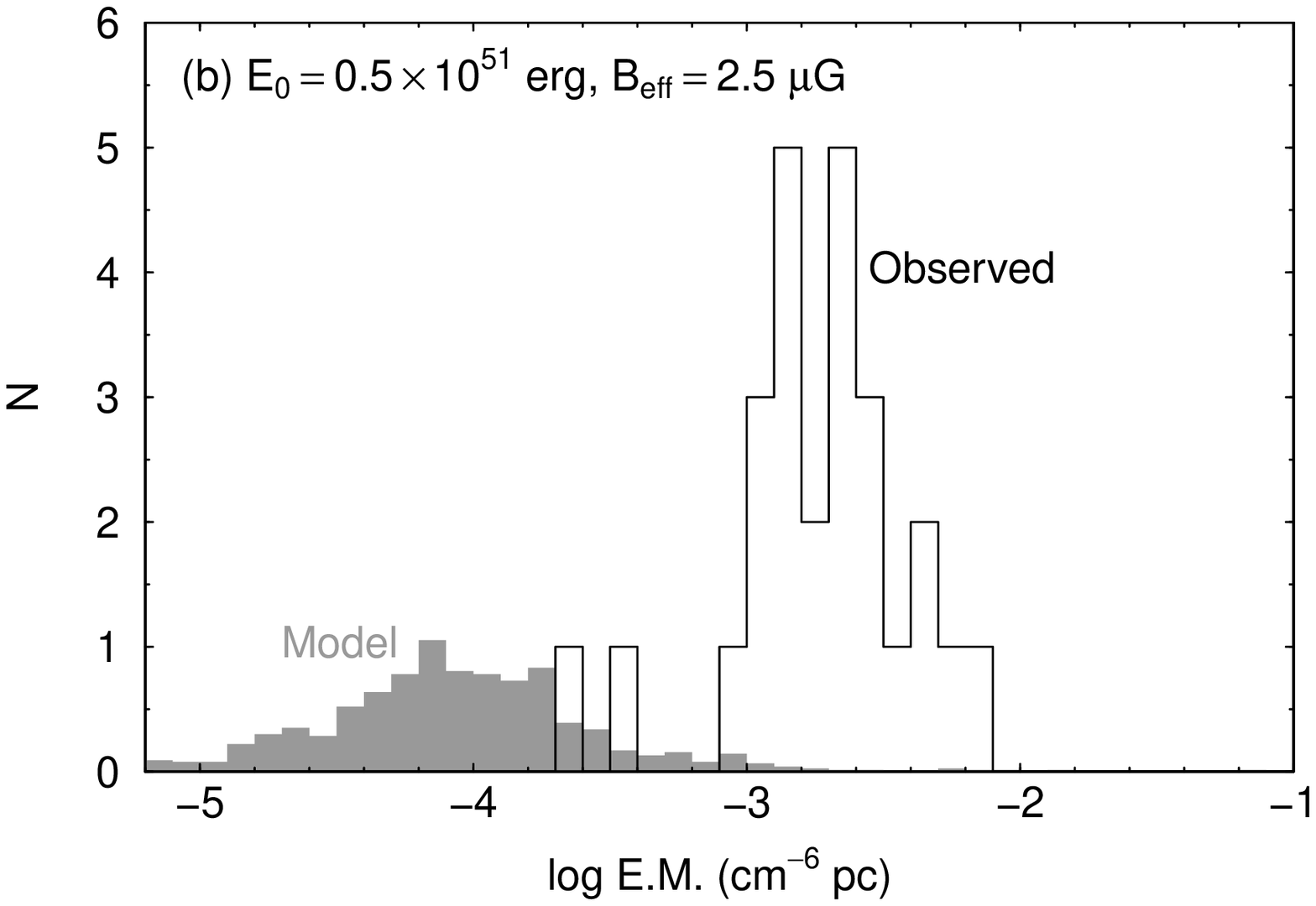} \\
\plotone{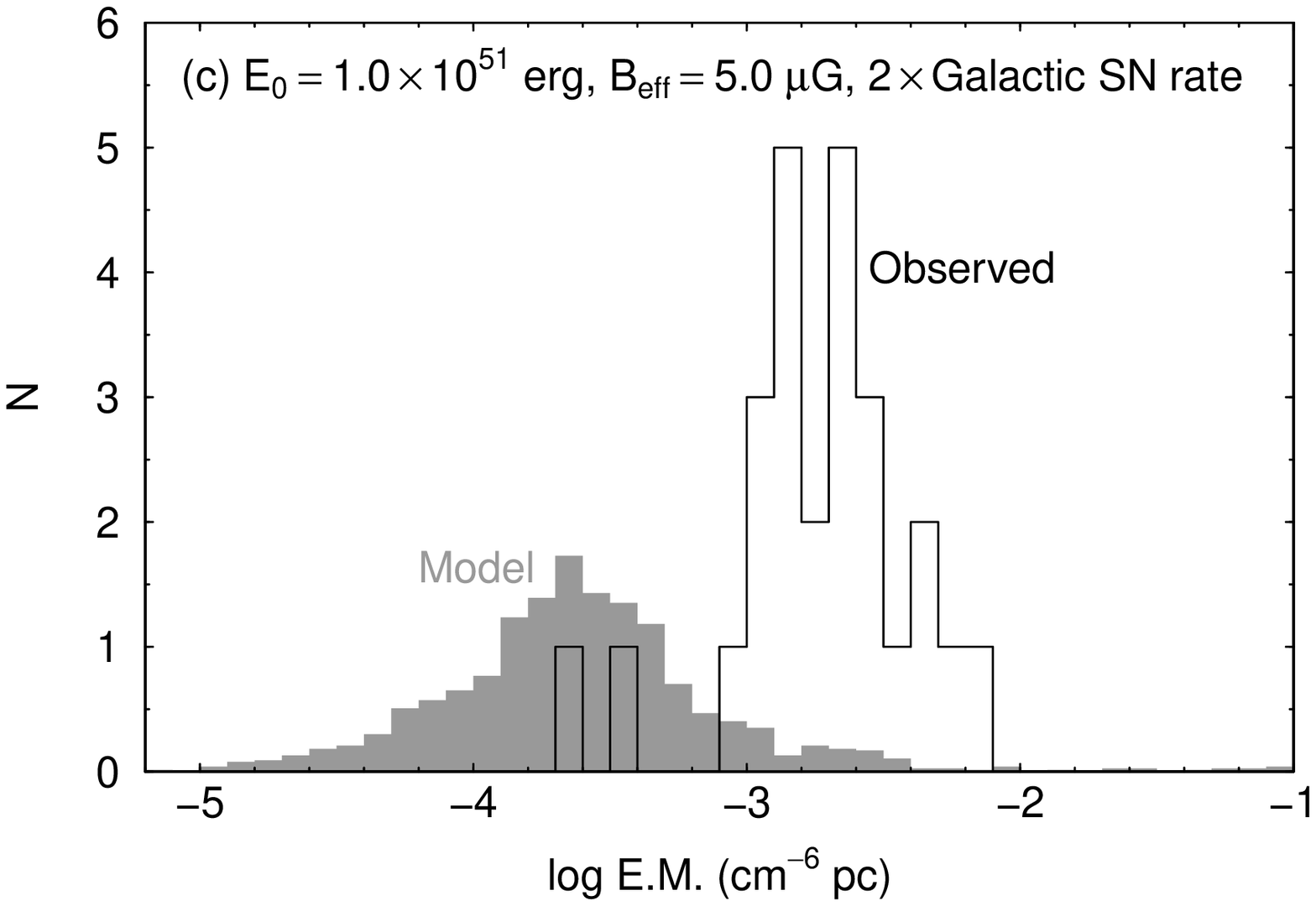}
\caption{Histograms comparing the halo X-ray temperatures (panel~a) and X-ray emission measures
  (panels~b and c) predicted by the extraplanar SNR model (solid gray) with those obtained from our
  \xmm\ observations (black outline). The observed emission measures have been multiplied by $\sin
  |b|$ (see text for details). The model temperatures and emission measures were derived by
  characterizing the model spectra with $1T$ models (see
  Section~\ref{subsubsec:CharacterizingSpectra}). Panels~(a) and (b) show model values derived from
  simulations with $E_0 = 0.5 \times 10^{51}~\erg$ and $\Beff = 2.5~\microgauss$, with a Galactic SN
  rate given by Equation~(\ref{eq:SNrate1}). Panel~(c) shows model emission measures derived from
  simulations with $E_0 = 1 \times 10^{51}~\erg$ and $\Beff = 5.0~\microgauss$, with a Galactic SN
  rate twice as large as that given by Equation~(\ref{eq:SNrate1}). The model histograms have been
  rescaled so they cover the same area as the corresponding observational histograms. Note that 64\%\
  of the model sightlines in panels~(a) and (b), and 42\%\ of the model sightlines in panel~(c) intercept
  no model remnants. Such sightlines have an undefined X-ray temperature and zero X-ray emission measure,
  and thus do not appear in the above plots.
  \label{fig:SNModel+Obs}}
\end{figure}

Figures~\ref{fig:SNModel+Obs}a and \ref{fig:SNModel+Obs}b show histograms comparing the halo X-ray
temperatures and emission measures predicted by the extraplanar SNR model with the corresponding
halo properties obtained from our \xmm\ observations. (Note that obs.~0406630201, which yielded $T =
11.2 \times 10^6~\K$, is not shown in Figure~\ref{fig:SNModel+Obs}a.)

The X-ray temperatures predicted by the model are in reasonably good agreement with the observed
temperatures. However, the emission predicted by this model is clearly too faint: the predicted X-ray emission measures are
typically an order of magnitude smaller than the observed values. Among the model sightlines that
intercept extraplanar SNRs, the median X-ray emission measure is $1.37^{+0.09}_{-0.12}$~dex smaller
than the median observed value of \EMsinb: $0.081 \times 10^{-3}$ versus $(1.90^{+0.41}_{-0.46})
\times 10^{-3}$~\emismeas.\footnote{Here and in Section~\ref{subsubsec:ISMModel+Obs}, the errors
  indicate the 90\%\ bootstrap confidence interval on the observed median.} As the predicted X-ray
temperatures are in reasonable agreement with the observed temperatures, the emission measure result
implies that the X-ray surface brightnesses are also typically underpredicted by an order of
magnitude. Note also that our Monte Carlo simulation predicts that about two-thirds of the
sightlines would intercept no extraplanar SNRs, whereas we observe hot halo gas on most, if not all,
of our sightlines.

\begin{deluxetable}{cccc}
\tablewidth{0pt}
\tablecaption{X-ray Emission Measures Predicted by Different Extraplanar SNR Models\label{tab:SNREM}}
\tablehead{
\colhead{$E_0$}                & \colhead{\Beff}                & \colhead{SN rate\tablenotemark{a}}   & \colhead{Median E.M.} \\
\colhead{($10^{51}~\erg$)}      & \colhead{(\microgauss)}        &                                      & \colhead{($10^{-3}~\emismeas$)}       \\
}
\startdata
0.5                            & 0                              & $1\times$                            & 0.028  \\
0.5                            & 2.5                            & $1\times$                            & 0.081\tablenotemark{b}  \\
0.5                            & 5.0                            & $1\times$                            & 0.19   \\
1.0                            & 5.0                            & $1\times$                            & 0.21   \\
1.0                            & 5.0                            & $2\times$                            & 0.23\tablenotemark{c}   \\
Observed                       &                                &                                      & $1.90^{+0.41}_{-0.46}$\tablenotemark{d}
\enddata
\tablecomments{The emission measures were obtained by fitting to the spectra above 0.4~\kev.}
\tablenotetext{a}{Relative to the rate given by Equation~(\ref{eq:SNrate1}).}
\tablenotetext{b}{Model used in Figures~\ref{fig:SNModel+Obs}a and \ref{fig:SNModel+Obs}b.}
\tablenotetext{c}{Model used in Figure~\ref{fig:SNModel+Obs}c.}
\tablenotetext{d}{Median of \EMsinb, with 90\%\ bootstrap confidence interval.}
\end{deluxetable}

The model values in Figures~\ref{fig:SNModel+Obs}a and \ref{fig:SNModel+Obs}b were calculated from
the SNR simulations in \citet{shelton06} with $E_0 = 0.5 \times 10^{51}~\erg$ and $\Beff =
2.5~\microgauss$. We find that increasing $E_0$, \Beff, or the assumed SN rate all increase the
predicted X-ray emission measures. Table~\ref{tab:SNREM} shows the median X-ray emission measures
predicted by extraplanar SNR models with different explosion energies, ambient magnetic fields, and
SN rates. As can be seen, increasing \Beff\ has the largest effect on the predicted X-ray emission
measures. However, it should be noted that the median model values in Table~\ref{tab:SNREM} are only
for the subset of sightlines that intercept at least one SNR. Increasing $E_0$ or the SN rate
increases the fraction of sightlines that intercept at least one SNR, while increasing
\Beff\ decreases that fraction.

Figure~\ref{fig:SNModel+Obs}c shows the histogram of X-ray emission measures predicted by the
\citet{shelton06} simulations with $E_0 = 1 \times 10^{51}~\erg$ and $\Beff = 5.0~\microgauss$, with
a Galactic SN rate that is twice that given by Equation~(\ref{eq:SNrate1}) (i.e., the 5th model in
Table~\ref{tab:SNREM}). The histogram of observed values is also plotted for comparison. The model
still significantly underpredicts the observed emission measures (it underpredicts the median value
by $0.91^{+0.09}_{-0.12}$~dex).  Furthermore, our new Monte Carlo simulation predicts that
$\sim$40\%\ of the sightlines would intercept no extraplanar SNRs -- we reiterate that we observe
hot halo gas on most, if not all, of our sightlines.
We will discuss the results presented here in Section~\ref{subsubsec:InSituSupernovaeDiscussion}.

\subsection{A Supernova-Driven Interstellar Medium}
\label{subsec:SNDrivenISM}

In this section, we examine another model in which the interstellar gas is heated by SN explosions.
This model is distinct from the previous model in a number of ways. The previous model considered
only isolated SNRs above $z = 130~\pc$, and the X-ray spectra were calculated from 1D hydrodynamical
simulations of individual remnants. Here, we use a 3D hyrodynamical simulation of vertically
stratified interstellar gas that is heated by discrete SN explosions \citep{joung06}. Unlike the previous model,
some SNe occur in clusters, the evolving SNRs can interact, older remnants may be re-energized by
new SNe, and SNe in the Galactic disk drive a fountain of hot gas up into the halo \citep{shapiro76,bregman80}.
However, ionization equilibrium was assumed, and magnetic fields neglected.

\subsubsection{Model Description}

The hydrodynamical simulation used here is described fully in \citet{joung06},
and the reader is referred to that paper for more details. The simulation was carried out using
Flash\footnote{Developed at the University of Chicago Center for Astrophysical Thermonuclear
  Flashes; http://flash.uchicago.edu/web/}, a parallelized Eulerian hydrodynamical code with
adaptive mesh refinement.  The simulation box extends from $z = -5~\kpc$ to $z = +5~\kpc$,
and has a size of $(0.5~\kpc)^2$ in the $xy$ plane.  The maximum spatial resolution is
$1.95~\pc$. The upper and lower boundaries have outflow boundary conditions, while the vertical sides
of the simulation box have periodic boundary conditions.

\begin{figure}
\centering
\includegraphics[width=0.3\linewidth]{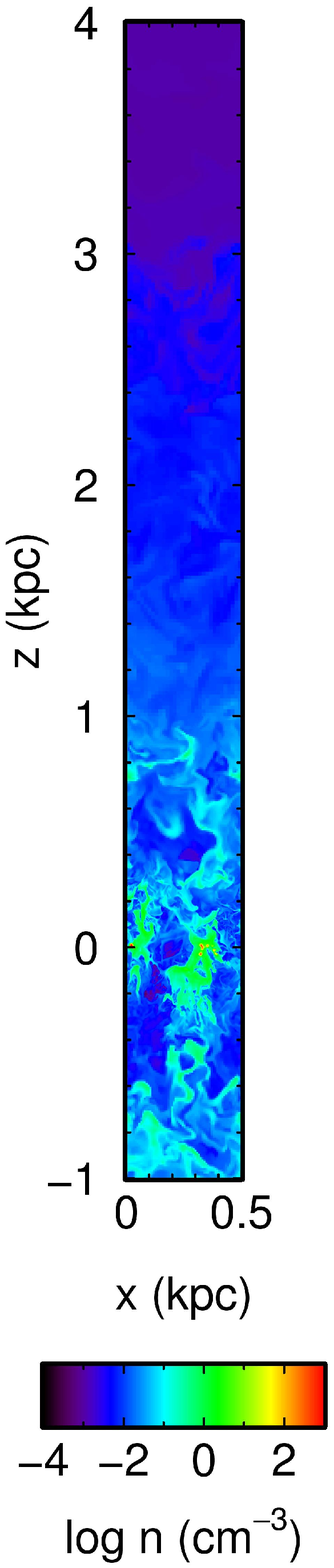}
\includegraphics[width=0.3\linewidth]{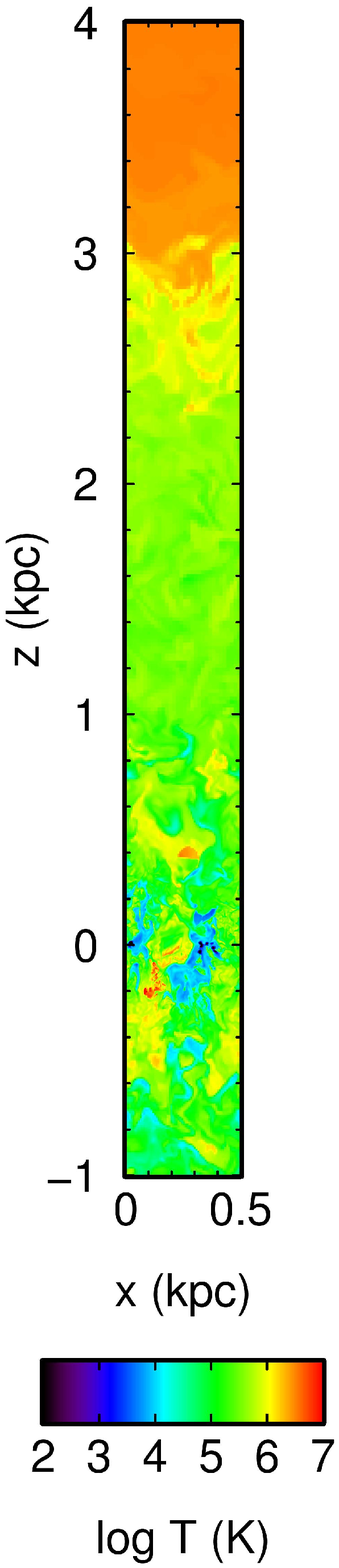}
\caption{Vertical slices of the density (left) and temperature (right) at $t = 120.0~\Myr$, from the
SN-driven ISM model of \citet{joung06}. Note that the full simulation box extends to $|z| = 5~\kpc$.
\label{fig:ISMDensity+Temperature}}
\end{figure}

The simulation box was initialized with $1.1 \times 10^4$~\K\ gas in hydrostatic equilibrium. This
gas was then heated by discrete SN explosions, each of which injected $10^{51}~\erg$ of energy into
a small region of the grid. These explosions generally occurred randomly in time and space, with
a rate
\begin{eqnarray}
        r(z) &=& r_\mathrm{Ia}(z) + r_\mathrm{II}(z)                                            \nonumber \\
             &=& ( 6.2 \e^{-|z|/325~\mathrm{pc}} + 167 \e^{-|z|/90~\mathrm{pc}} )~\kpc^{-3}~\Myr^{-1},~~~~
\label{eq:SNrate2}
\end{eqnarray}
although 3/5 of the Type~II SNe occurred in clusters of 7 to $\approx$40
explosions. The above rates differ from those in Equation~(\ref{eq:SNrate1}) in two ways. Firstly,
\citet{joung06} assumed higher Galactic SN rates than \citet{ferriere98a}: $1/330~\yr^{-1}$ versus
$1/445~\yr^{-1}$ (Type~I) and $1/44~\yr^{-1}$ versus $1/52~\yr^{-1}$ (Type~II). Secondly,
Equation~(\ref{eq:SNrate1}) considers only isolated Type~II SNe, whose average height is larger than
the average height of all Type~II progenitors (266 versus 90~\pc; \citealt{ferriere95}). The model
also includes diffuse heating, due to photoelectric emission from UV-irradiated dust grains, and
radiative cooling. Figure~\ref{fig:ISMDensity+Temperature} shows vertical slices of the density and
temperature at $t = 120.0~\Myr$.

We calculated X-ray spectra for 242 sightlines looking vertically upward and downward from the
Galactic midplane. The viewpoints of these sightlines formed an $11 \times 11$ grid in the $xy$
plane, with a grid spacing of $\approx$49~\pc. The spectra were calculated using the
\citet{raymond77} spectral code (updated by J.~C. Raymond \& B.~W. Smith, 1993, private
communication with R.~J. Edgar), assuming that the material along the line of sight is in CIE.
We assumed that the gas on the grid is optically thin. We ignored the emission from the
first 100~\pc\ of each sightline, as such material is not in the Galactic halo. In our observational
analysis, emission from within $\sim$100~\pc\ of the midplane is attributed to our foreground model
component, derived from \rosat\ shadowing data.

\begin{figure}
\centering
\plotone{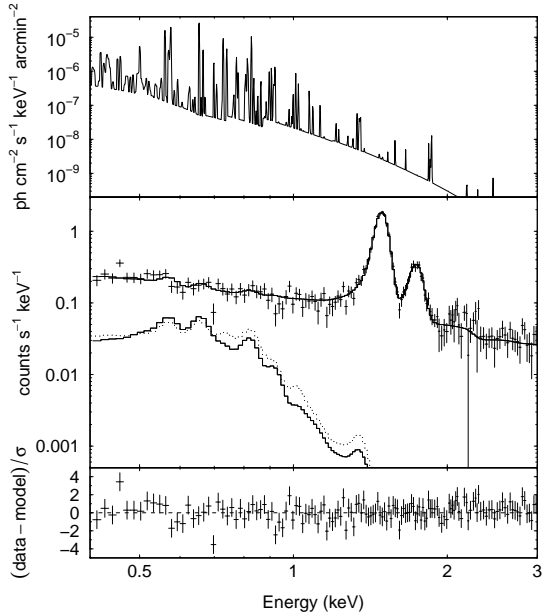}
\caption{As Figure~\ref{fig:FakeSNRSpectrum}, but using the SN-driven ISM model as the input halo
  model.
  \label{fig:FakeISMSpectrum}}
\end{figure}

As with the previous model, we used the method described in
Section~\ref{subsubsec:CharacterizingSpectra} to characterize the model spectra with $1T$ models,
and used the resulting X-ray temperatures and X-ray emission measures to compare with our
observations. As in Section~\ref{subsubsec:CharacterizingSpectra}, our simulated \xmm\ spectra
included foreground, extragalactic background, and instrumental background components, as well as the
halo emission from the SN-driven ISM model. The emission from the halo and extragalactic components
was subjected to absorption with $\NH = 1.7 \times 10^{20}~\pcmsq$. This column density was also
used in the subsequent spectral fitting, from which we obtained the characteristic X-ray
temperatures and X-ray emission measures. Figure~\ref{fig:FakeISMSpectrum} shows a simulated
\xmm\ spectrum for one of our model sightlines, along with the best-fitting model. The differential
emission measure for the input halo spectrum is shown in Figure~\ref{fig:DEM} (dashed line). Because
we assume that the plasma is in CIE, only plasma with $T \ga 10^6~\K$ will contribute in the
\xmm\ band. Although the input model predicts emission from a range of temperatures, a model with a
$1T$ halo generally fits the simulated spectra well.

It should be noted that our model spectra do not take into account absorption by cold gas on the
grid -- when we characterized our model spectra, we assumed that the absorbing column ($\NH = 1.7
\times 10^{20}~\pcmsq$) was located entirely beneath the hot X-ray--emitting gas. For $\sim$2/3 of
the model sightlines, the column density of cold gas that is mixed in with the bulk of the
X-ray-emitting gas is $<$$10^{19}~\pcmsq$ (i.e., an order of magnitude smaller than the column
densities used in our \xmm\ analysis). We have investigated the effect of ignoring on-grid
absorption by creating simulated \xmm\ spectra from an XSPEC model of the form $\mathtt{phabs} \ast
(\mathtt{raymond} + \mathtt{phabs} \ast \mathtt{raymond})$, and characterizing the resulting spectra
with a model of the form $\mathtt{phabs} \ast \mathtt{raymond}$. As with our previous simulations,
the column density of the first \texttt{phabs} component was fixed at $1.7 \times 10^{20}~\pcmsq$.
The two \texttt{raymond} components in the input model had the same emission measure, and
temperatures of $1.5 \times 10^6~\K$ and $2.5 \times 10^6~\K$ (it does not matter which component is
the hotter -- our conclusion is the same either way). We found that the characteristic X-ray
temperatures and X-ray emission measures obtained were not strongly affected by the column density
between the two \texttt{raymond} components in the input model, at least up to column densities of a
$\mbox{few} \times 10^{20}~\pcmsq$. We therefore conclude that ignoring on-grid absorption will not
adversely affect our results.

In the following, we also compare the predicted X-ray surface brightnesses with our
observations. These values were extracted directly from the model spectra.

\subsubsection{Comparing the SN-Driven ISM Model with Observations}
\label{subsubsec:ISMModel+Obs}

\begin{figure}
\centering
\plotone{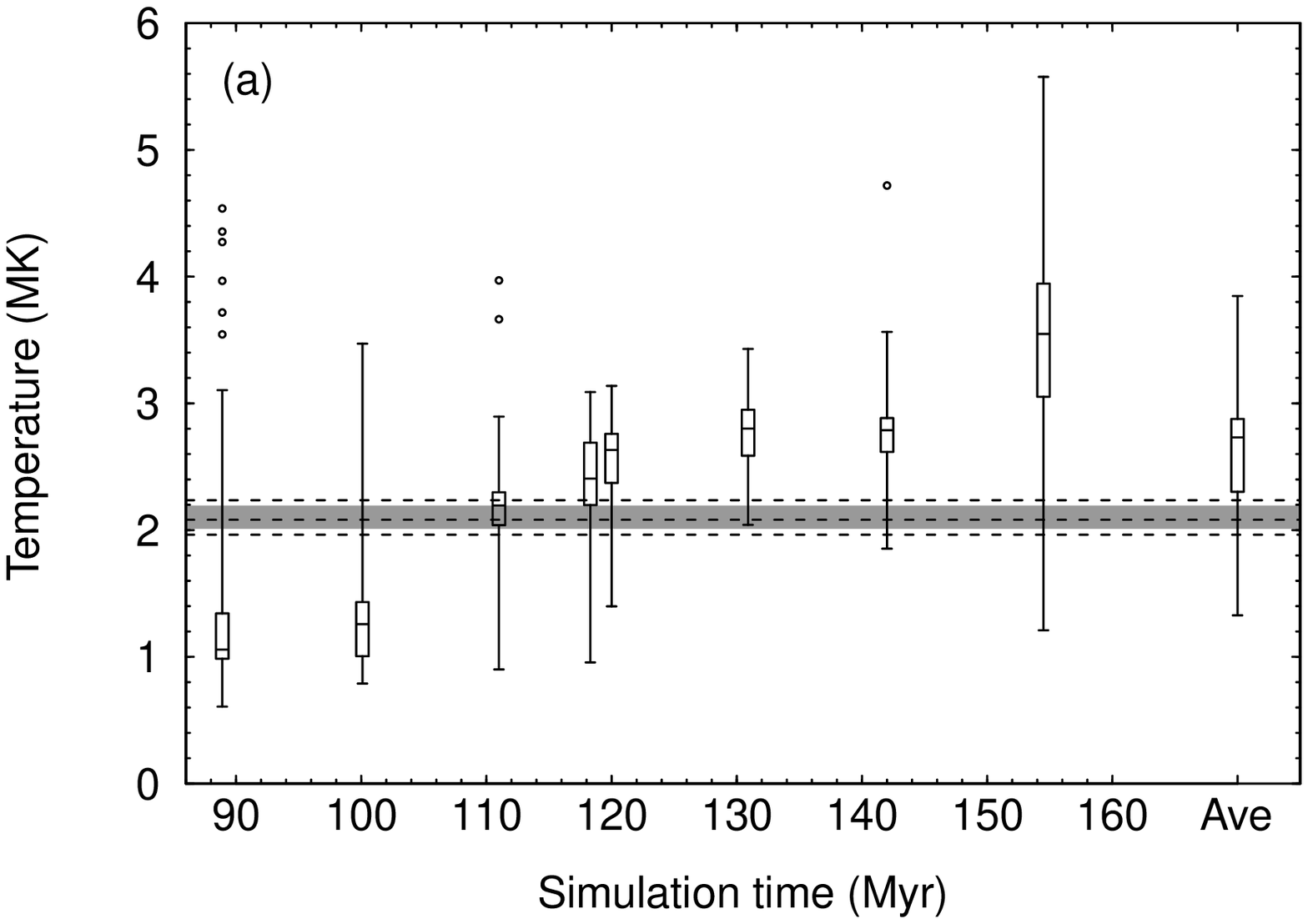}  \\
\plotone{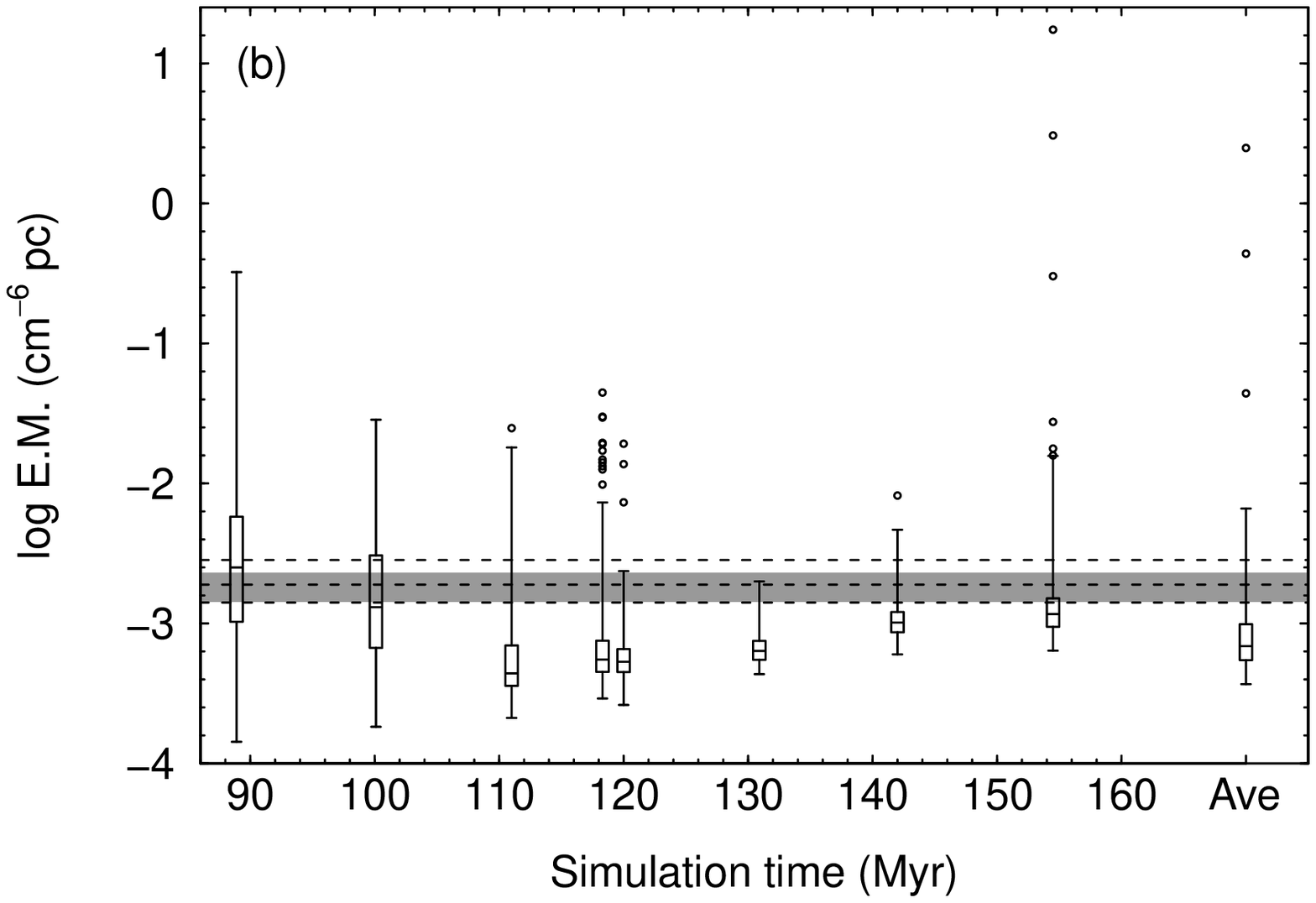} \\
\plotone{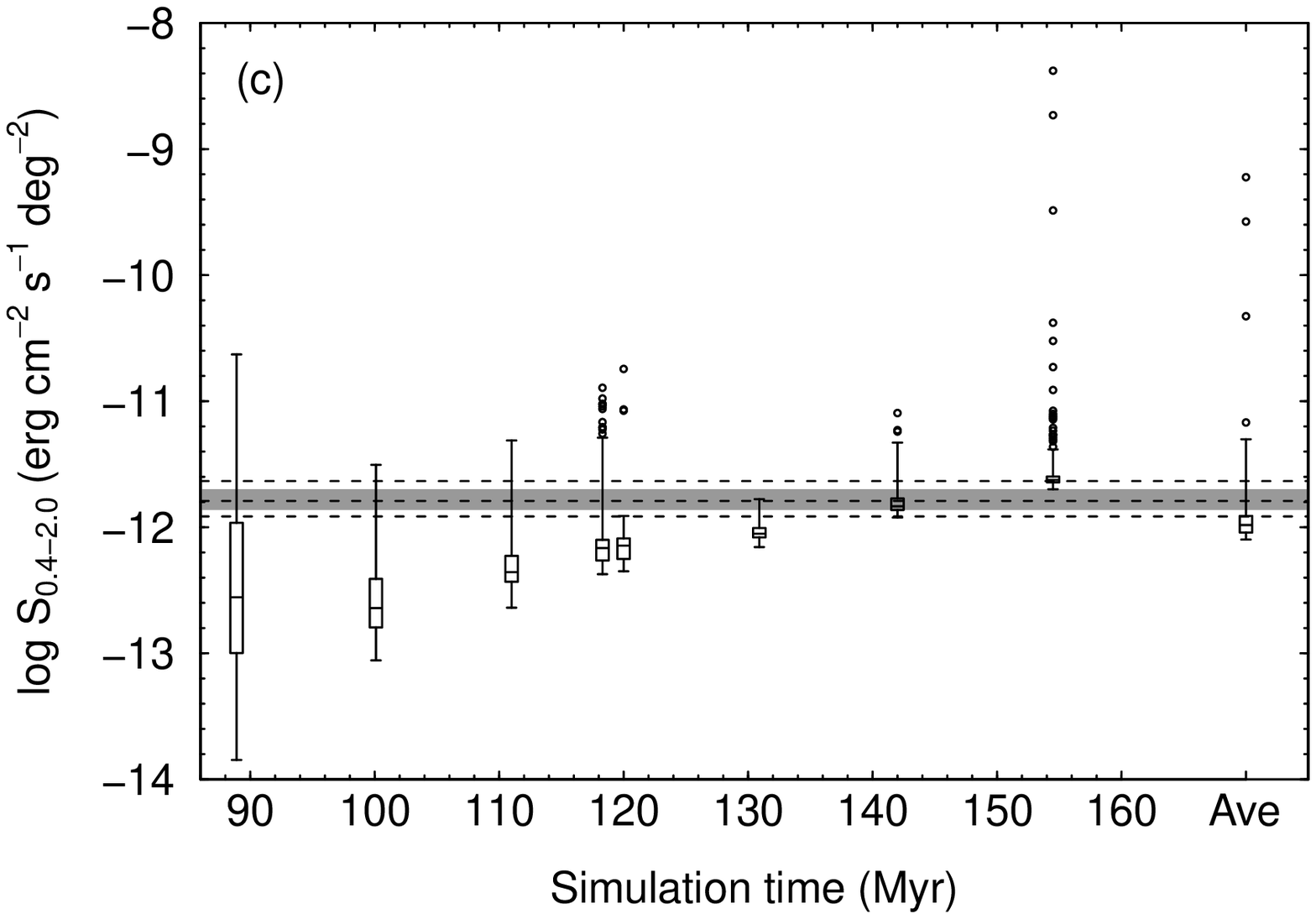}
\caption{Comparison of the halo (a) X-ray temeratures, (b) X-ray emission measures, and (c)
  intrinsic 0.4-2.0~\kev\ surface brightness predicted by the SN-driven ISM model with the values
  obtained from our \xmm\ observations. The boxplots show the predicted values for a range of
  simulation times. The rightmost datapoint shows the results obtained by averaging the spectra
  from several of the timesteps (see text for details). Each box indicates the median and quartiles,
  and the whiskers extend to the most extreme point that is no more than 5 times the interquartile
  range from the box; any outliers are plotted individually with open circles. The horizontal dashed
  lines indicate the observed median and quartiles, and the gray band indicates the 90\%\ bootstrap
  confidence interval on the observed median.
  \label{fig:ISMModel+ObsBoxplot}}
\end{figure}

Figure~\ref{fig:ISMModel+ObsBoxplot} shows the time variation of the X-ray spectral properties
predicted by the SN-driven ISM model. The model has not settled down to a steady state -- the
typical X-ray temperature and X-ray surface brightness rise steadily throughout the period
shown. However, in this temperature regime, an increase in X-ray surface brightness can be brought
about by an increase in temperature as well as by an increase in emission measure. The
typical X-ray emission measure does not rise steadily throughout the plotted period.

Figure~\ref{fig:ISMModel+ObsBoxplot} also compares the predicted X-ray properties with the observed values
from our \xmm\ analysis. Similarly to Section~\ref{subsubsec:SNRModel+Obs}, we multiplied our
observed emission measures and surface brightnesses by $\sin |b|$ before comparing them with the
model values. At the earlier times, the model halo is too cool and too faint in the
0.4--2.0~\kev\ band.  Around $t = 110~\Myr$, the predicted X-ray temperatures are in reasonable
agreement with the observed value, but the halo is typically a factor of $\sim$4 too faint. At later
times the model halo is too hot; by $t = 155~\Myr$ it is also somewhat brighter than is observed,
although the median predicted surface brightness is within 50\%\ of the observed median.

From Figure~\ref{fig:ISMModel+ObsBoxplot} is it not clear what state the halo is tending toward.
\Citet{avillez04} estimated that a steady state halo in a simulation such as this should be reached
in $\la$180~\Myr. However, it is possible that, instead of settling down to a steady state that is
hotter and brighter than the observed halo, the variation in X-ray temperature and surface
brightness is part of a slow oscillation about a mean state, with a period of at least $\sim$130~\Myr.
Determining the final state predicted by this model would probably require running it for at least
several more tens of megayears, which is not currently practical. We therefore try two different approaches
in comparing the simulation data to our observations.

\begin{figure*}
\centering
\plottwo{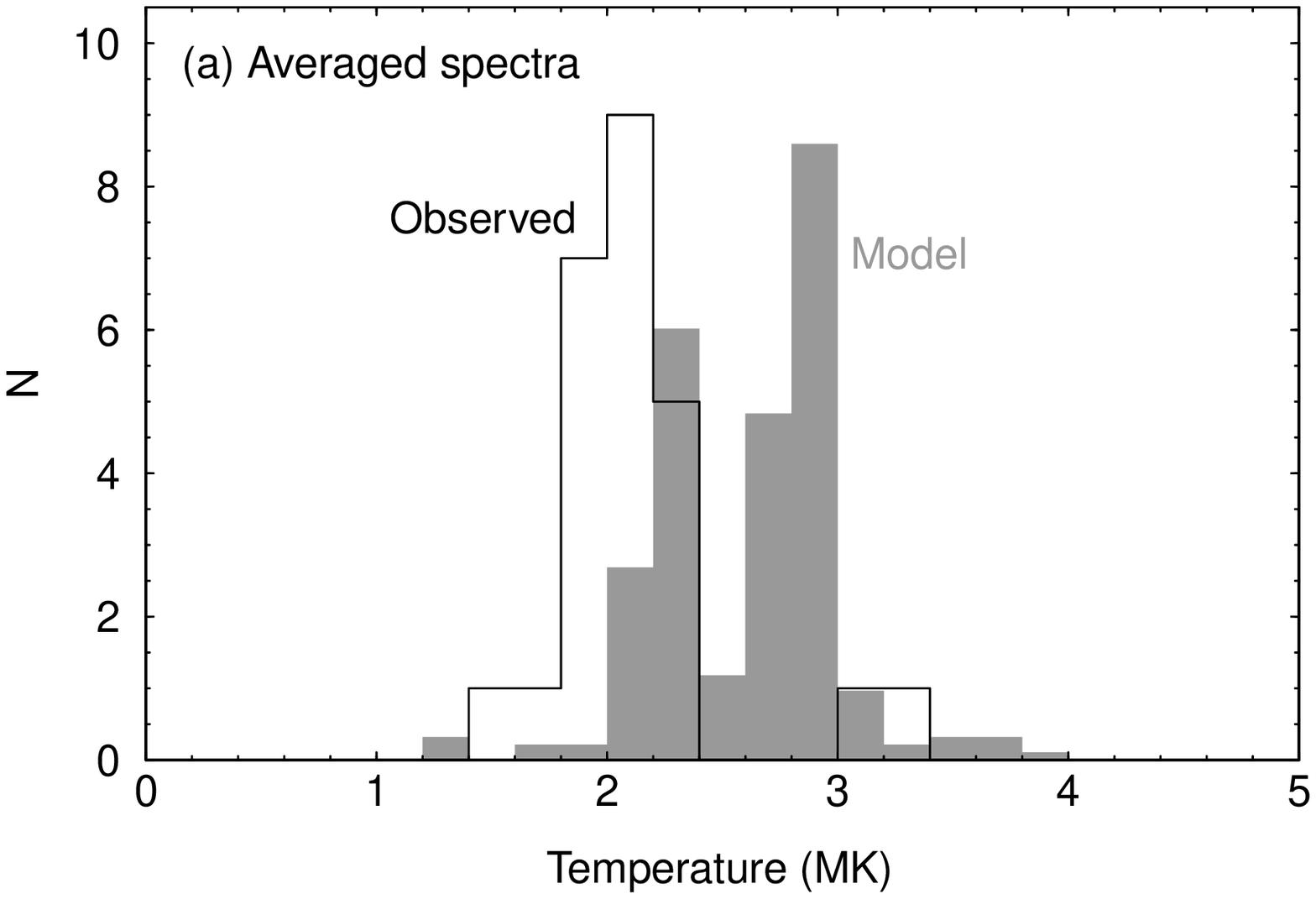}{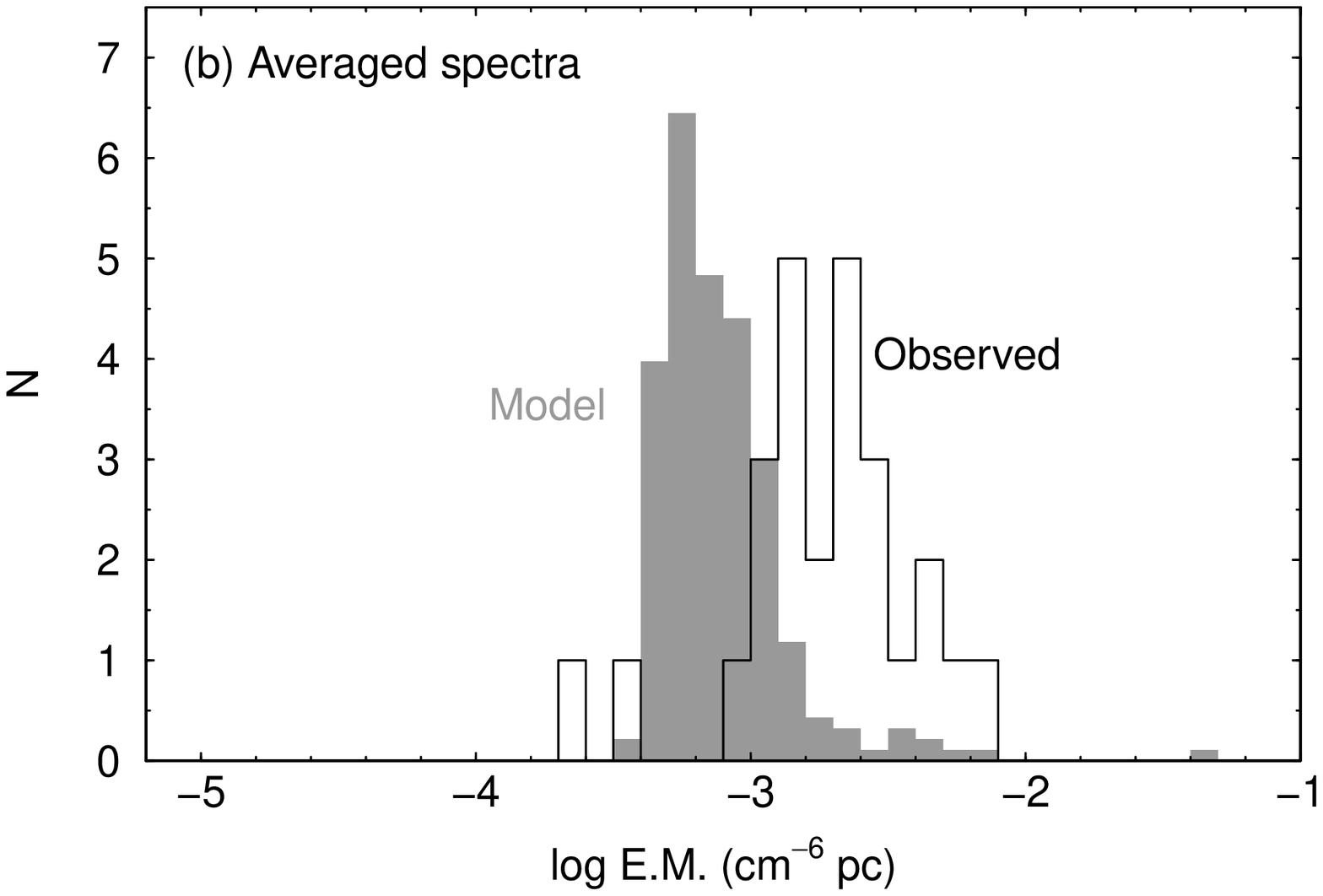} \\
\plottwo{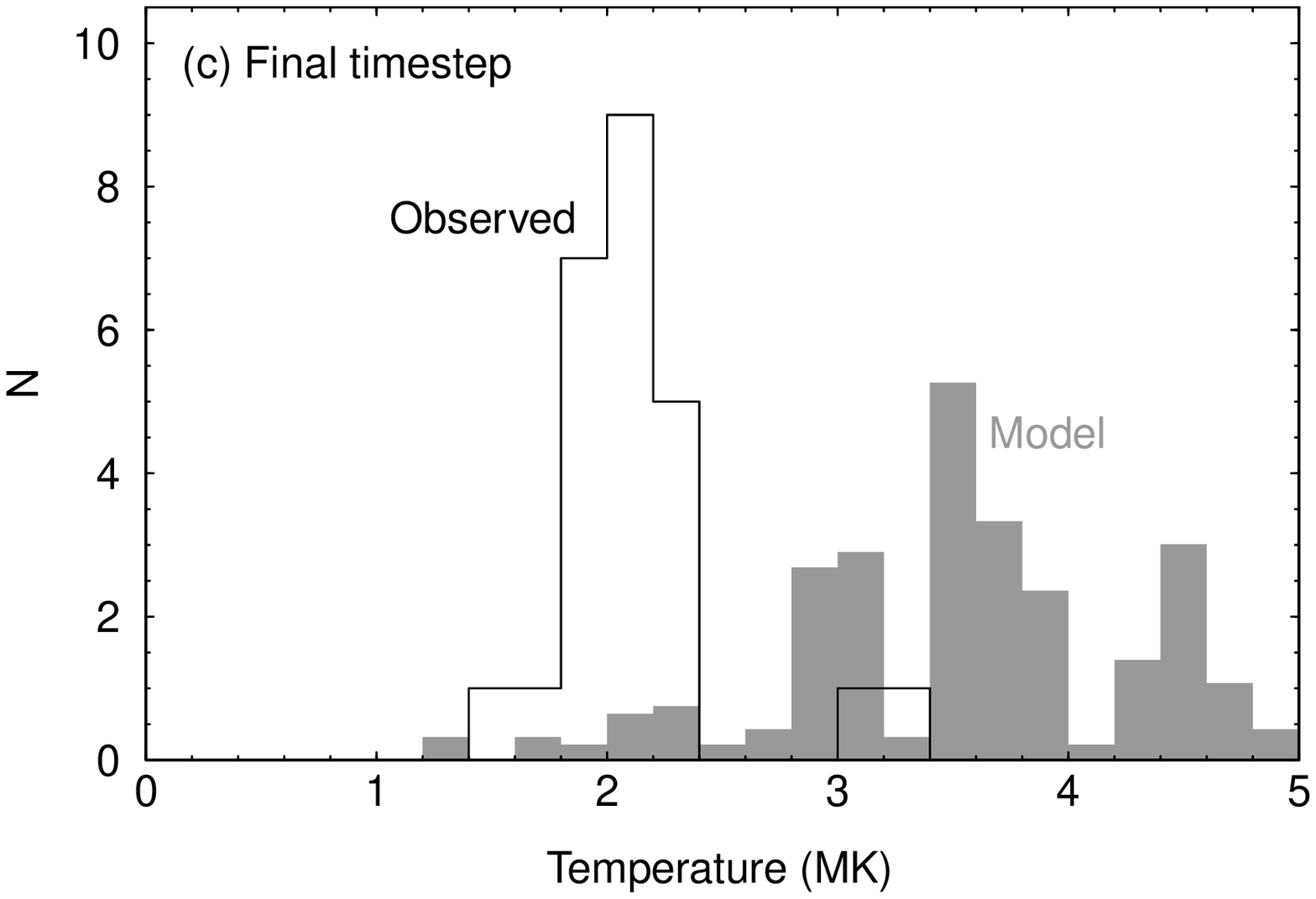}{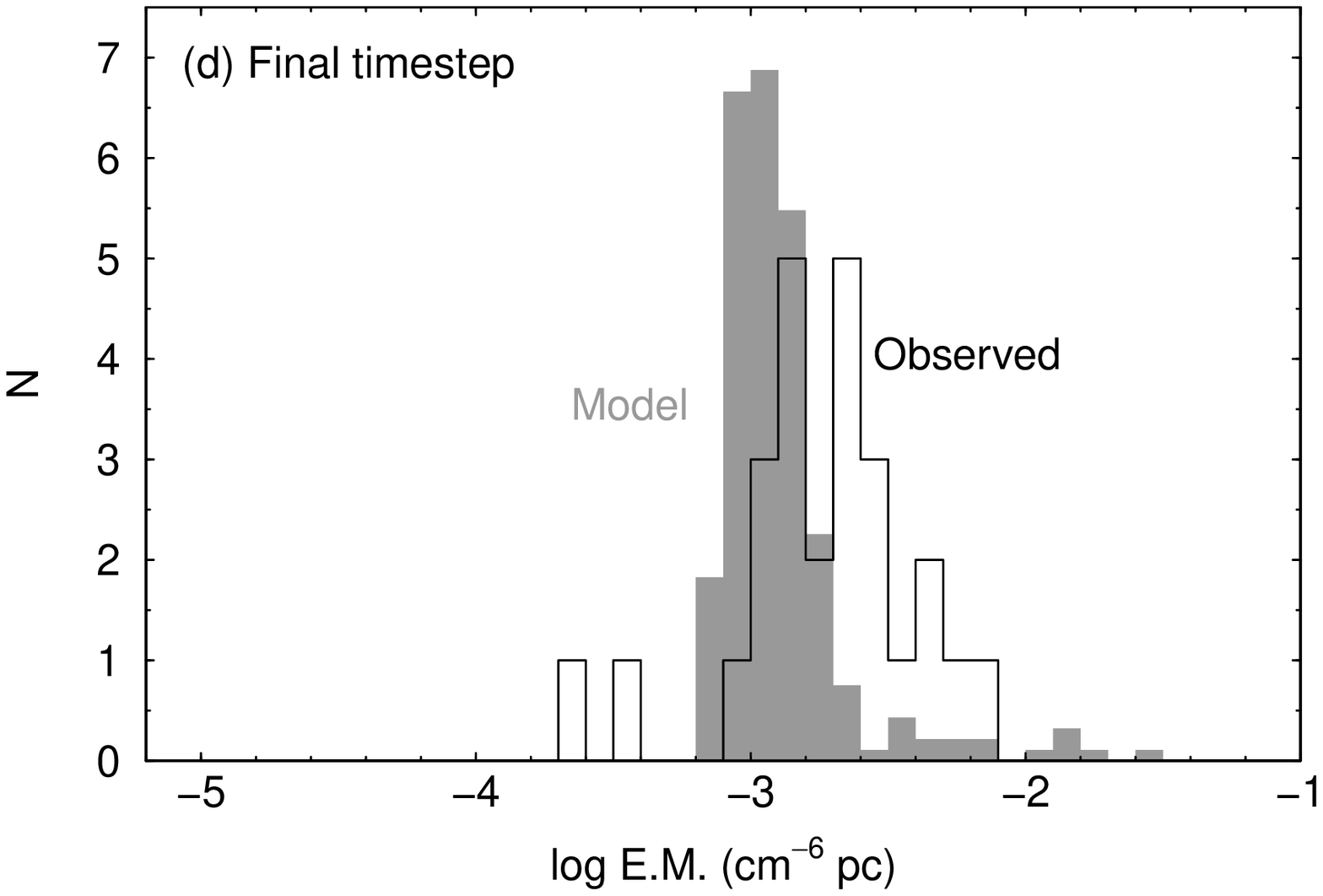}
\caption{Histograms comparing the halo X-ray temperatures (panel~a and c) and X-ray emission
  measures (panels~b and d) predicted by the SN-driven ISM model (solid gray) with those obtained
  from our \xmm\ observations (black outline). The observed emission measures have been multiplied
  by $\sin |b|$ (see Section~\ref{subsubsec:SNRModel+Obs}). The model temperatures and emission
  measures were derived by characterizing the model spectra with $1T$ models (see
  Section~\ref{subsubsec:CharacterizingSpectra}). The model histograms have been rescaled so they
  cover the same area as the corresponding observational histograms. Panels (a) and (b) show the
  results obtained by averaging the spectra from several model timesteps in
  Figure~\ref{fig:ISMModel+ObsBoxplot} (see text for details), while panels (c) and (d) show the
  predictions from the final timestep ($t = 155~\Myr$) in Figure~\ref{fig:ISMModel+ObsBoxplot}.
  \label{fig:ISMModel+ObsHistogram}}
\end{figure*}

Our first approach is to assume that the data shown in Figure~\ref{fig:ISMModel+ObsBoxplot}
represent roughly half a cycle in the oscillation of the halo about a mean state. The simulation
models only a narrow column of the halo. As our observation directions are not all toward the
Galactic poles, our sightlines sample different spatial locations in the halo, which, in a
statistical sense, should correspond to different times throughout the cycle. We therefore proceed
by averaging the spectra for each sightline from 7 of the 8 timesteps shown in
Figure~\ref{fig:ISMModel+ObsBoxplot} (we don't use the data from $t = 118$~\Myr, so as not to
oversample the times around $t \sim 120~\Myr$). We characterize the resulting averaged spectra with
$1T$ models, using the procedure described previously. The results are compared with our
observations in Figure~\ref{fig:ISMModel+ObsBoxplot} (the rightmost datapoint, labeled ``Ave''), and
in Figures~\ref{fig:ISMModel+ObsHistogram}a and \ref{fig:ISMModel+ObsHistogram}b.

Our second approach is to assume that the final time in Figure~\ref{fig:ISMModel+ObsBoxplot} ($t =
155~\Myr$) gives our best estimate of the steady state that the halo is approaching. We use the
predictions from that final time in Figures~\ref{fig:ISMModel+ObsHistogram}c and
\ref{fig:ISMModel+ObsHistogram}d.

Whether we use the averaged spectra or the final timestep, the predicted halo temperature is too
high. The median predicted temperatures are $2.73 \times 10^6$~K (averaged spectra) and $3.55 \times
10^6$~K (final timestep), against an observed median of $(2.08^{+0.11}_{-0.07}) \times 10^6$~K. The
averaged spectra underpredict the halo emission measure, although the discrepancy is not as large as
with the extraplanar SNR model. The median predicted X-ray emission measure is
$0.44^{+0.09}_{-0.12}$~dex smaller than the median observed value of \EMsinb: $0.69 \times 10^{-3}$
versus $(1.90^{+0.42}_{-0.46}) \times 10^{-3}$~\emismeas. For comparison, the discrepancy for the
extraplanar SNR model is $1.37^{+0.09}_{-0.12}$~dex. For the spectra from the final timestep, the
median predicted X-ray emission measure ($1.17 \times 10^{-3}~\emismeas$) is in better agreement
with the observed median (the discrepancy is $0.21^{+0.09}_{-0.12}$~dex). The predicted
0.4--2.0~\kev\ surface brightnesses are generally in reasonably good agreement with the observed
values, although there are a few sightlines in the final timestep of the model that exhibit much
greater surface brightnesses. The median predicted values are within 50\%\ of the median observed
value of $\Stotal \sin |b|$: $1.04 \times 10^{-12}$ (averaged spectra) and $2.37 \times 10^{-12}$
(final timestep), versus $(1.62^{+0.30}_{-0.05}) \times 10^{-12}$~\flux\ \pdegsq\ (observed). We
will discuss the results presented here in Section~\ref{subsubsec:SNDrivenISMDiscussion}.

\section{DISCUSSION}
\label{sec:Discussion}

\subsection{The Temperature and Emission Measure of the Galactic Halo}
\label{subsec:HaloTandSB}

In Section~\ref{subsec:ModelDiscussion}, below, we will discuss the various physical models of the
hot halo that we examined in the previous section. Here, we discuss our \xmm\ halo measurements,
and compare the results with those from other recent studies.

Our measured halo temperatures are typically $\sim$$2 \times 10^6$~K -- the median temperature is
$2.08 \times 10^6$~K, and the lower and upper quartiles are $1.96 \times 10^6$ and $2.24 \times
10^6$~K, respectively. The halo temperatures measured here, using a thermal plasma model, are more
tightly constrained than those inferred from the \OVII/\OVIII\ intensity ratio (Paper~I). This is
partly because fitting with a thermal plasma model uses more of the information in an observed
spectrum, and partly because the errors on the intensities of individual lines are not combined in
the temperature uncertainty.

\begin{figure}
\centering
\plotone{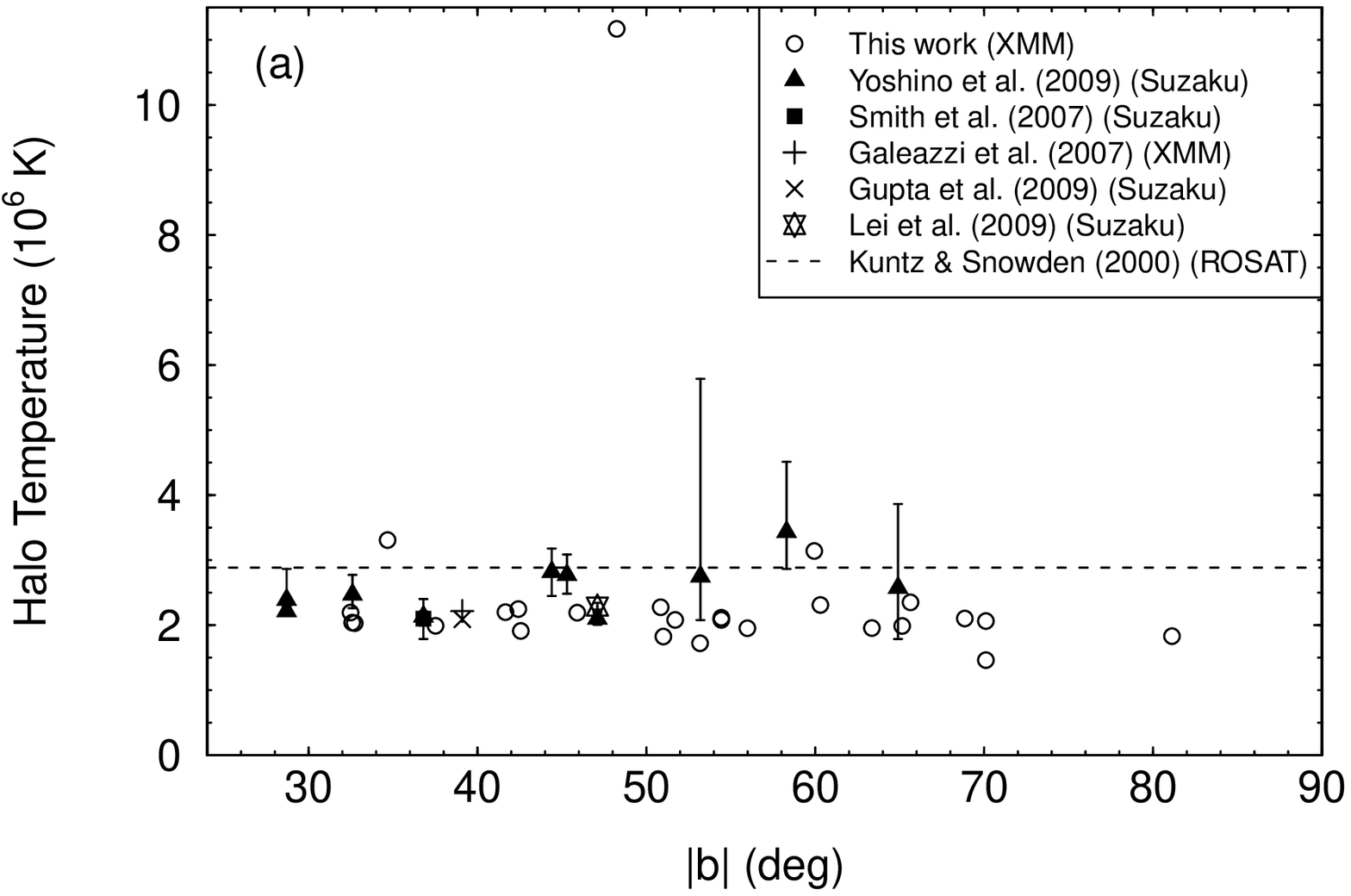} \\
\plotone{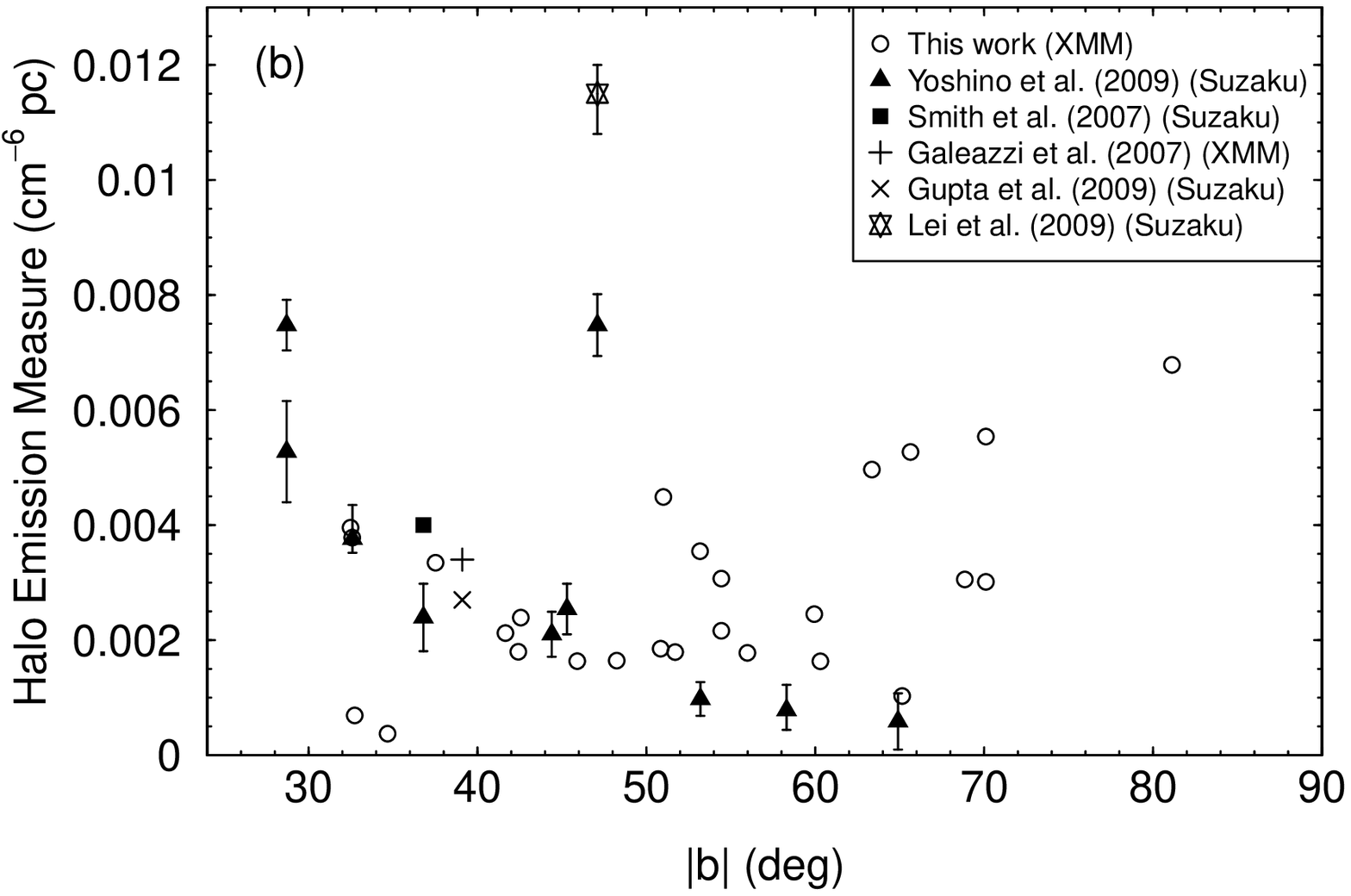}
\caption{Halo (a) temperatures and (b) emission measures plotted against Galactic latitude. The open
  circles show the results from this work, with the error bars omitted for clarity.  The solid
  triangles show the results from \citet{yoshino09} with $|b| > 20\degr$; specifically, the results
  obtained with a fixed foreground model for all sightlines (their Table~6).  The solid square shows
  the results from a \suzaku\ shadowing observation of MBM~12 \citep{smith07a}. The ``$+$'' and the
  ``$\times$'' show the results from shadowing observations of MBM~20, carried out with \xmm\ and
  \suzaku, respectively \citep{galeazzi07,gupta09b}. Specifically, the \xmm\ result is the result
  obtained with frozen abundances from Table~2 in \citet{galeazzi07}, and the \suzaku\ result is
  from row~3 of Table~2 in \citet{gupta09b}. The star shows the results from a \suzaku\ shadowing
  observation of a dusty filament in the southern Galactic hemisphere \citep{lei09}; specifically,
  model~2 from their Table~1. The horizontal dashed line shows the temperature derived by
  \citet{kuntz00} from the \rosat\ All-Sky Survey.
  \label{fig:OtherObs}}
\end{figure}

Figure~\ref{fig:OtherObs}a compares our halo temperatures with various other measurements made with
\rosat, \xmm, and \suzaku\ \citep{kuntz00,smith07a,galeazzi07,yoshino09,lei09,gupta09b}. For papers
with multiple results, the figure caption indicates the specific results that we have plotted. Our
temperatures are typically lower than the temperature obtained by \citet{kuntz00}, using
\rosat\ All-Sky Survey data. The most likely reason for this is that \citet{kuntz00} used a
two-temperature ($2T$) model for the halo, whereas as we used a $1T$ model. An additional,
lower-temperature halo component could account for some of the lower-energy flux in the \xmm\ band,
enabling the other component to shift toward higher temperatures. However, in practice we found that
we were unable to constrain a $2T$ model using our \xmm\ data, as \xmm's sensitivity does not extend
to as low energies as \rosat's.  A $1T$ halo model adequately described our \xmm\ spectra.

Our halo temperatures are generally in good agreement with those measured by \citet{smith07a},
\citet{galeazzi07}, \citet{gupta09b}, \citet{lei09}, and \citet{yoshino09}. All of these
measurements are from $1T$ halo models. Note that the \citet{yoshino09} temperatures were obtained
with a fixed foreground model for all their sightlines, and Fe and Ne abundances (relative to O)
that were free to vary (their Table~6). When they fitted their model with an independent foreground
model and fixed abundances for each sightline, \citeauthor{yoshino09} obtained systematically higher
halo temperatures (their Tables~3 and 4). However, \citeauthor{yoshino09} argue that these higher
temperatures are mostly determined by the Fe-L and Ne-K emission, and are inconsistent with the
behavior of the \OVII\ and \OVIII\ emission. When they measured the halo temperatures that we have
used in Figure~\ref{fig:OtherObs}a, \citeauthor{yoshino09} found that about half of their sightlines
required either an overabundance of Fe and Ne, or an additional hotter emission component.

The halo temperatures measured here and in other recent studies are fairly constant across the sky.
This is most likely because gas with $T \la 1 \times 10^6~\K$ would be difficult to detect with
\xmm\ (unless it has a large emission measure), while gas with $T \ga 3 \times 10^6~\K$ would escape
from the Galactic potential well \citep[e.g.,][]{bregman09}

Figure~\ref{fig:OtherObs}b compares our halo emission measures with those from the studies discussed
above. In general, our values are in good agreement with the values from the other studies. The large
emission measure obtained by \citet{lei09} may be partly due to their using \citet{wilms00} abundances.
The \citet{wilms00} oxygen abundance is a factor of 1.8 smaller than the \citet{anders89} value. As oxygen
emission tends to dominate the halo X-ray spectrum, a smaller oxygen abundance will result in a larger
X-ray emission measure.

There is considerable scatter in the halo emission measure, with the values spanning an order of
magnitude ($\sim$0.0005--0.006~\emismeas). This patchiness to the halo has already been pointed out by
\citet{yoshino09} and in Paper~I. We will discuss in more detail the various physical models that we
have examined in Section~\ref{subsec:ModelDiscussion}, below, but we note here that a patchy halo
favors an inhomogeneous, stochastic heat source, such as SNe, as opposed to accretion of extragalactic material,
which we would expect to be fairly homogeneous.

\subsection{Physical Models of the Hot Halo}
\label{subsec:ModelDiscussion}

Here, we discuss each of the models presented in Section~\ref{sec:ComparisonWithModels} in turn.

\subsubsection{The Disk Galaxy Formation Model}
\label{subsubsec:DiskGalaxyFormationDiscussion}

In Section~\ref{subsec:DiskGalaxyFormation} we examined a disk galaxy formation model
\citep{toft02,rasmussen09}, which predicts the existence of a hot halo extended over tens
of kiloparsecs, formed from material falling into the galaxy's potential well. Using results from
\citet{rasmussen09}, we find that the emission-weighted mean temperature predicted for a Milky
Way-sized galaxy ($\vc \sim 220~\kmps$) is in good agreement with the median observed halo
temperature, but the model underpredicts the X-ray luminosity of the halo by at least an order of
magnitude.

The above results suggest that the extended hot halo predicted by disk galaxy formation models is
not a major contributor to the halo X-ray emission that we observe with \xmm. However, this
interpretation is complicated by the simulations of \citet{crain10}. They argue that the stellar
feedback in the \citet{rasmussen09} simulations is too weak, resulting in too much mass ending up in
stars and too little mass in hot gas. Ideally, we would compare the surface brightness of the
accreted extragalactic material in the \citet{crain10} simulations with the halo surface brightness
obtained from our \xmm\ observations, but this is not possible for two reasons. Firstly, in addition
to infalling extragalactic material, the \citeauthor{crain10} model includes interstellar gas that
was heated by stars and SNe and transferred upwards by an outflow (similar to the galactic fountain
in the model described in Section~\ref{subsec:SNDrivenISM}). Secondly, their model predictions are
in terms of luminosity rather than surface brightness. Because gas heated by stars and SNe will tend
to be concentrated nearer the disk, compared with the more extended halo of accreted extragalactic
material, it may provide a much larger surface brightness per unit luminosity than the extended halo
gas. Predictions of the separate contributions to the surface brightness due to the gas heated by
stellar processes and the extended halo are not currently available from the \citeauthor{crain10}
model. Such predictions are needed to determine whether or not accreted extragalactic material makes
a major contribution to the observed halo X-ray emission.

Although surface brightness predictions are not currently available, we can make rough comparisons
between the total X-ray luminosity predicted by \citet{crain10} and the halo luminosity expected
from the \xmm\ data. An \LX--$T$ plot derived from their simulations shows that galaxies with
emission-weighted halo temperatures of $\sim$$2 \times 10^6~\K$ have 0.3--2.0~\kev\ luminosities of
$\sim$3--$40 \times 10^{39}~\ergps$ (R. Crain, 2010, private communication), compared with $<$$0.2
\times 10^{39}~\ergps$ from \citet{rasmussen09}. In Section~\ref{subsec:DiskGalaxyFormation} we
derived a Galactic halo luminosity of $1.9 \times 10^{39}~\ergps$ from our observations, which is
slightly smaller than the range of luminosities predicted by \citet{crain10}. However, this observed
luminosity was calculated assuming that the observed emission came from a uniform sphere of radius
$\Rsph = 15~\kpc$. Figure~2 in \citet{crain10} implies that the emission in their model is much more
extended than this, with $\Rsph \sim 50~\kpc$. Using this larger radius leads to an observed halo
luminosity of $1.3 \times 10^{40}~\ergps$, which lies within the range predicted by
\citeauthor{crain10}.

\subsubsection{The Extraplanar SNR Model}
\label{subsubsec:InSituSupernovaeDiscussion}

In Section~\ref{subsec:InSituSupernovae} we examined a model in which the hot halo gas is contained
in an ensemble of isolated extraplanar SNRs \citep{shelton06}. The X-ray temperatures predicted by
this model are in good agreement with the observed values. However, the predicted X-ray emission
measures, obtained from model spectra above 0.4~\kev, are typically an order of magnitude smaller
than the observed values. We can increase the predicted X-ray emission measures by increasing the SN
explosion energy, $E_0$, the effective ambient magnetic field, \Beff, or the SN rate. However,
doubling all three of these parameters still resulted in X-ray emission measures that were too
small.

Of the above three parameters, \Beff\ has the largest effect on the predicted X-ray emission
measures (see Table~\ref{tab:SNREM}). Increasing \Beff\ increases the non-thermal pressure, and
therefore increases the compression of the hot X-ray--producing bubble. This increased compression
increases the density and temperature of the bubble, increasing the X-ray emission measure inferred
from the predicted emission in the \xmm\ band. We have not carried out simulations with $\Beff >
5.0~\microgauss$, but if we extrapolate the values in Table~\ref{tab:SNREM}, we find we would need a
\Beff\ of a few tens of microgauss to match the observed emission measures. Such a magnetic field
implies a non-thermal pressure $\Pnt/k \ga 10^5~\presalt$. This is an implausibly high non-thermal
pressure for the halo, as it is several times larger than the midplane value
\citep{boulares90,ferriere01}.  In addition, the model with $\Beff = 2.5~\microgauss$ predicts that hot gas
would be seen on only $\sim$$1/3$ of sightlines. As noted above, increasing \Beff\ results in
smaller, hotter remnants.  In addition, these smaller remnants are brighter and so shorter-lived. As a
result, increasing \Beff\ means that even fewer sightlines would intercept remnants. In reality, we
find hot gas on most, if not all, of our \xmm\ sightlines.  Therefore, we cannot bring the
extraplanar SNR model into agreement with our observations by increasing the assumed effective
ambient magnetic field.

Increasing the SN rate increases the number of SNRs expected to lie along a given sightline, and so
increases the predicted X-ray emission measures. However, we would have to increase the SN rate
given by Equation~(\ref{eq:SNrate1}) by a factor of $\sim$6 in order to give the same increase in the
median X-ray emission measure that we see when we increase \Beff\ from 2.5 to 5.0~\microgauss. To
match the observations, we would have to increase the SN rate by an even larger factor. The Galactic
SN rate of $\sim$2 per century is constrained to within a factor of $\sim$2 (see the online
Supplementary
Information\footnote{http://www.nature.com/nature/journal/v439/n7072/suppinfo/nature04364.html}
for \citealp{diehl06}). If the SN scale heights in Equation~(\ref{eq:SNrate1}) are well constrained,
the halo SN rate is also constrained to within a factor of $\sim$2. Even if we double the SN scale
heights in Equation~(\ref{eq:SNrate1}), the integrated SN rate above $|z| = 130~\pc$ only increases by
a factor of 2.5. Therefore, to bring the extraplanar SNR model into agreement with our observations
we would have to increase the SN rate by an unrealistic amount. We also cannot bring the model into
agreement with our observations by increasing $E_0$, as Table~\ref{tab:SNREM} shows that the median
X-ray emission measure depends only weakly on $E_0$.

We therefore conclude that the hot halo gas that we observe with \xmm\ is not primarily due to an
ensemble of isolated extraplanar SNRs. Most of this population of remnants would be relatively old
($\mathrm{age} > 1$~\Myr) and faint in the \xmm\ band. It should be noted, however, that this result
does not imply that extraplanar SNRs do not contribute to the hot halo gas at all. In the \xmm\ band
(above 0.4~\kev), their contribution is masked by brighter emission from an SN-driven galactic
fountain (see Section~\ref{subsubsec:SNDrivenISMDiscussion}, below). At lower energies, extraplanar
SNRs could still contribute significantly to the hot gas observed in the 1/4~\kev\ band, as
originally suggested by \citet{shelton06}.  In addition, young, bright remnants, although rare, do
produce emission that is detectable by \suzaku\ \citep{henley09} and that should also be detectable
by \xmm.

\subsubsection{The SN-Driven ISM Model}
\label{subsubsec:SNDrivenISMDiscussion}

In Section~\ref{subsec:SNDrivenISM} we examined a hydrodynamical simulation of vertically stratified
interstellar gas, driven by SN explosions \citep{joung06}. Unlike the extraplanar SNR model, the
SNRs do not evolve in isolation, and the model results in a galactic fountain of hot gas up into the
halo \citep{shapiro76,bregman80}. Also, for this model we assumed that the gas is in CIE, whereas the extraplanar SNR model
includes self-consistent modeling of the ionization evolution.

As with the extraplanar SNR model, we folded the model spectra through the \xmm\ response, added
photon noise, and characterized the resulting spectra with $1T$ models. Before going on to discuss
the comparison of the resulting X-ray temperatures and emission measures with our
\xmm\ observations, we shall look at how well these X-ray spectral properties reflect the properties
of the gas on the hydrodynamical grid.

\begin{figure}
\centering
\plotone{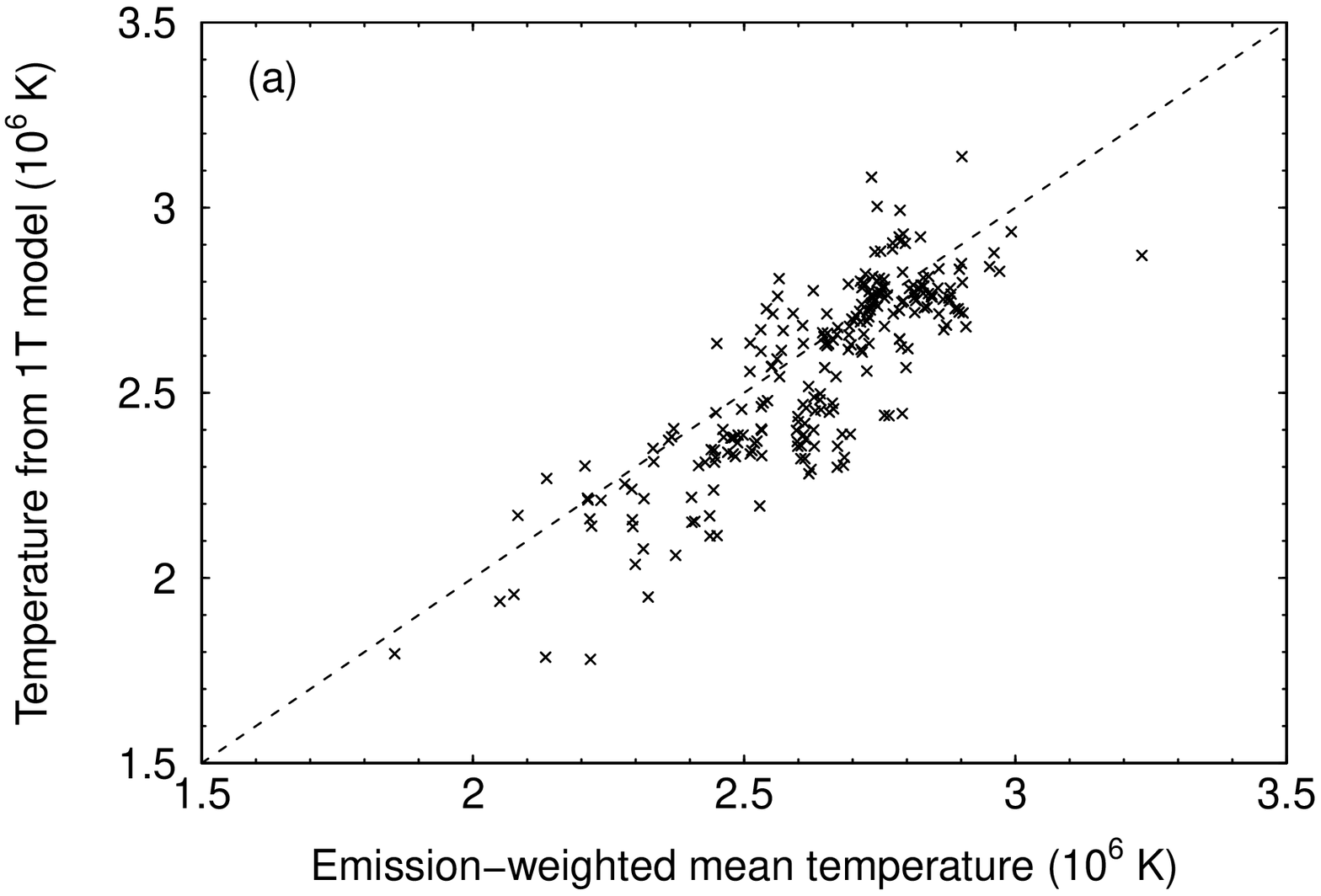} \\
\plotone{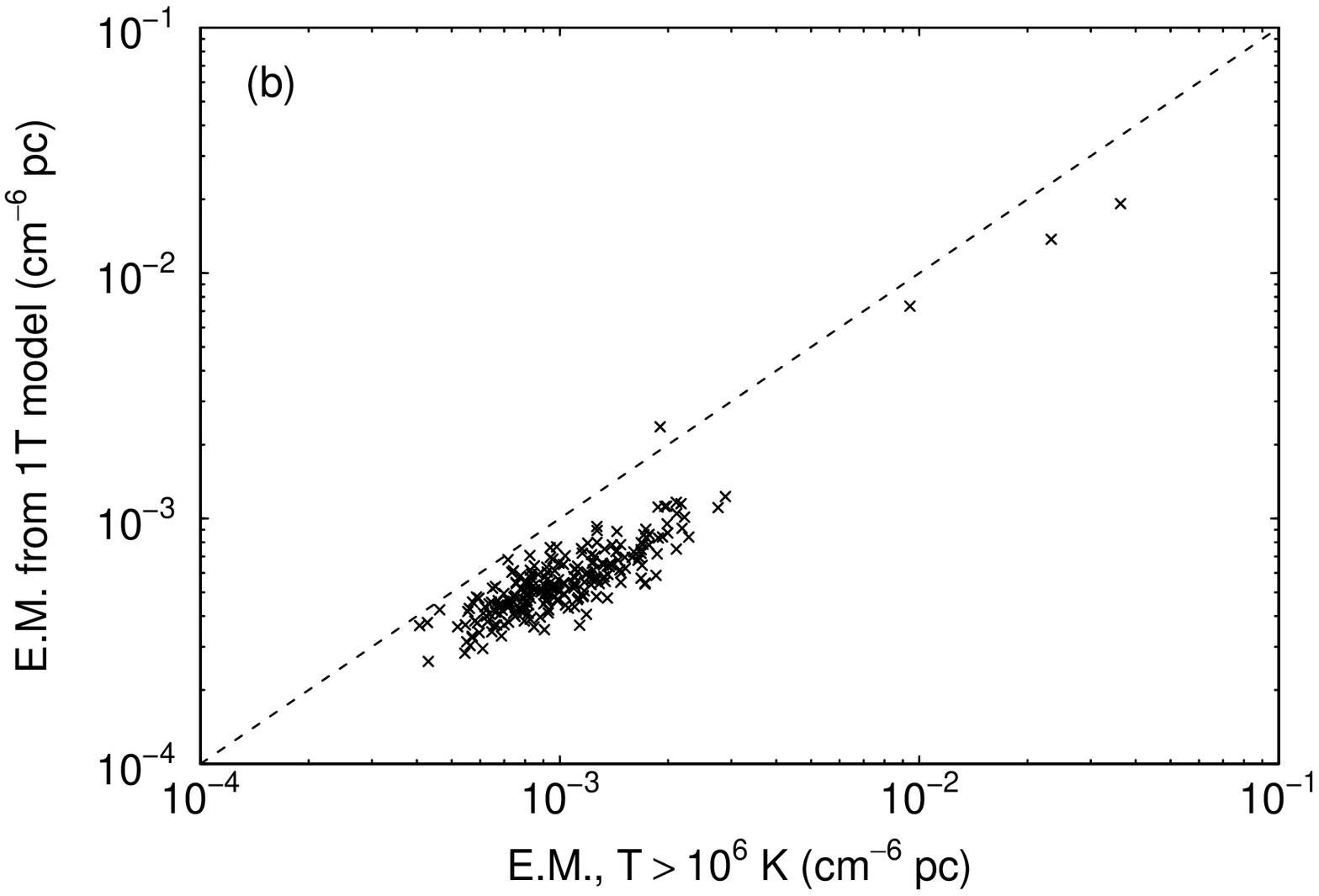}
\caption{Comparison of X-ray spectral properties derived from $1T$ models (ordinates) and properties
  derived directly from the hydrodynamical data (abscissae) for each sightline through the
  SN-driven ISM model at $t = 120~\Myr$.
  \label{fig:1T-vs-model}}
\end{figure}

Figure~\ref{fig:1T-vs-model}a compares the X-ray temperatures derived from the $1T$ models with the
mean temperature along each model sightline through the hydrodynamical grid, weighted by the
0.4--2.0~\kev\ emission. We find that the X-ray temperatures are in reasonable agreement with the
emission-weighted mean temperatures (typically within $0.4 \times 10^6~\K$).
Figure~\ref{fig:1T-vs-model}b compares the X-ray emission
measure for each sightline with the emission measure of the gas with $T > 10^6~\K$ along that
sightline (we use this quantity because there is no obvious analog to the emission-weighted mean
temperature). The X-ray emission measure underestimates the emission measure of gas with $T >
10^6~\K$, although the two quantities are well correlated and generally agree within a factor of
2.5. Overall, we conclude that the properties derived from the $1T$ models are good
characterizations of the hot gas on the hydrodynamical grid

The 0.4--2.0~\kev\ halo surface brightness predicted by this model is in good agreement with the
observations (within $\sim$50\%). As the emission predicted by this model is much brighter, it will
mask the contribution from extraplanar SNRs. Note that this model does include some \textit{in situ}
heating by extraplanar SNe. The integrated halo SN rate above $|z| \sim 100~\pc$ is $\sim$2 times
larger in this model than in the extraplanar SNR model (although the volumetric SN rate falls off
more rapidly with $|z|$ in this model). However, this larger integrated SN rate is not enough to
explain the differences in X-ray surface brightness between the models: this model predicts emission an order of
magnitude brighter than the extraplanar SNR model. The majority of the halo X-ray emission in this
model comes from hot gas that is driven from the disk into the halo by a galactic fountain. As
mentioned in Section~\ref{subsubsec:DiskGalaxyFormationDiscussion}, it is possible that the emission
from a galactic outflow or fountain also masks the contribution from an extended hot halo of
accreted material.

Although the median predicted X-ray surface brightness and X-ray emission measure are in good
agreement with the observed medians,
the halo X-ray temperature predicted by this model is $\sim$$1 \times 10^6~\K$ larger than
our measured value, independent of whether the simulated halo is undergoing a long-period ($P \ga
100~\Myr$) oscillation about some mean state, or is settling down to a steady state. Given the
uncertainties in the model predictions (due to the time variability shown in
Figure~\ref{fig:ISMModel+ObsBoxplot}), these discrepancies are probably insufficient to rule out
this model. Nevertheless, we discuss below three possible causes for the temperature discrepancy:
(1) the simulation overpredicts the temperature of the hot gas in the halo; (2) the temperature of
the hot gas does not accurately predict the X-ray spectrum, because the gas is not in ionization
equilibrium; or (3) a bias in our spectral analysis causes us to underestimate the observed halo
X-ray temperature. Correcting for whichever of these effects turn out to be important may also
affect the X-ray surface brightnesses and emission measures. However, it is not easy to foresee
whether these corrections will increase or decrease the discrepancies between the predicted and
observed surface brightnesses and emission measures.

The first possible cause of the temperature discrepancy is that the simulation overpredicts the
temperature of the hot gas in the halo. The simulation does not include thermal conduction, which
could in principle lower the temperature of the hot X-ray-emitting gas, although thermal conduction
in the ISM can be suppressed by magnetic fields. \Citet{avillez04} argue that turbulent diffusion is
more efficient at mixing hot and cold gas than thermal conduction. However, the efficiency of this
mixing will be underestimated if the mixing is not fully resolved. In the simulation used here, the
spatial resolution is between 1.95 and 15.6~\pc, but the adaptive mesh refinement criterion is
chosen to focus maximum numerical resolution on the 200~\pc\ above and below the midplane, so
the gas is less well resolved in the halo than in the disk. As a result, it is possible that
turbulent mixing of hot and cold gas is under-resolved in the halo, resulting in an overestimate of
the temperature of the X-ray-emitting gas \citep{fujita09}.

A lack of spatial resolution may also suppress the radiative cooling of the hot gas -- averaging the
hot gas density over large cells eliminates local denser regions that would radiate more efficiently
\citep{avillez04}. \Citet{avillez04} found that the filling factor of hot ($T > 10^{5.5}~\K$) gas
was not significantly affected by the spatial resolution of their simulations. However, there is
insufficient information to determine how the emission-weighted mean temperature of the
X-ray-emitting gas would be affected. Simulations with higher spatial resolution in the halo
will help determine whether or not the mixing of hot and cold gas and the radiative cooling of hot
gas are adequately resolved for predicting the X-ray emission.

\citet{joung06} pointed out that the average gas density at several disk scale heights and beyond
($0.2~\kpc \la |z| \la 2.5~\kpc$) is somewhat underpredicted in their model compared to
observations.  They suggested that additional components of pressure from the magnetic field and
cosmic rays may contribute significantly to the support in the vertical direction (see Section~3.1
of their paper).  Additional vertical pressure support will lead to larger disk scale heights and
hence larger gas masses and lower temperatures at 1--2~\kpc\ heights, which will reduce the
discrepancy between the observations and the SN-driven ISM model.  Preliminary results from
magnetohydrodynamics simulations are in agreement with this expectation (A. Hill et~al., 2010, in
preparation).

The second possible cause for the discrepancy between the predicted and observed X-ray temperatures
is that the hot gas is out of equilibrium. This gas was shock-heated by SNe -- this rapid heating
causes the ionization temperature (which determines the X-ray spectrum) to lag behind the kinetic
temperature (which is the quantity obtained from the hydrodynamical simulation).  This ionization
evolution was followed self-consistently in the 1D SNR models described in
Section~\ref{subsec:InSituSupernovae}, but not in this 3D simulation. CIE is reached on a timescale
of $t_\mathrm{eq} \sim 10^{12} \Ne^{-1}$~\s, where \Ne\ is the electron density in
\pcc\ \citep{masai94}. In the simulation, the density of gas with $T > 10^6~\K$ is
$\la$$10^{-3}~\pcc$, implying $t_\mathrm{eq} \ga 30$~\Myr. This is similar to the dynamical
timescale ($5~\kpc / c_\mathrm{s} \sim 30$~\Myr, where $c_\mathrm{s}$ is the sound speed in
$10^6~\K$ gas), and so we would expect the hot gas to be at least partially underionized.

Calculating the X-ray spectrum of an underionized plasma requires knowledge of the ionization
balance, which in turn depends on the history of the plasma. However, we note that an underionized
plasma will have fewer higher-stage ions (e.g., O$^{+7}$) relative to lower-stage ions (e.g.,
O$^{+6}$) than we would expect from the kinetic temperature. Therefore, if the halo is underionized,
modeling the observed emission with a CIE model will underestimate the kinetic temperature of the
halo gas. We have confirmed this by simulating \xmm\ observations of an underionized plasma (using
the XSPEC \texttt{nei} model), and characterizing the resulting simulated spectra with $1T$ CIE
models (for simplicity, here we just use the \texttt{nei} model as an input for our simulations,
instead of the full multicomponent model used in Section~\ref{subsubsec:CharacterizingSpectra}).
For $\Ne t \la 10^9$~\pcc\ \s, CIE models do not fit the simulated spectra well (these models
underestimate the low-energy flux).  However, for $\Ne t \sim 10^9$--$10^{12}$~\pcc\ \s, $1T$ CIE
models give good fits to the simulated spectra, but underestimate the input kinetic temperature.

The above discussion considers only the hot gas being underionized. An additional possible source of
X-rays is delayed recombination from overionized gas -- gas that has undergone rapid adiabatic
cooling as it expands into the halo \citep{breitschwerdt94,breitschwerdt99}. With our current
simulation data, it is difficult to estimate the contribution of delayed recombination to the halo
emission, relative to the emission from the hot gas. Simulations that can track the ionization
evolution of the plasma are needed to determine the extent to which the assumption of CIE affects
the predicted spectra, and hence the derived X-ray temperature.

The third possible cause of the temperature discrepancy is that something (other than the assumption
of CIE) is biasing the temperatures measured from our \xmm\ observations. The discrepancy is not
because we characterize the halo emission with a $1T$ model, as opposed to a $2T$ model (e.g.,
\citealt{kuntz00}; see Section~\ref{subsec:HaloTandSB}), because we characterize the model spectra
in the same way. If we are underestimating the halo temperature, the most likely cause is that our
foreground (LB and/or SWCX) model is too faint; in particular, that it underestimates the foreground
\OVII\ emission (underestimating the foreground \OVII\ emission means that the halo \OVII\ emission
is overestimated relative to the halo \OVIII\ emission, causing the measured halo temperature to
shift to a lower value). However, our halo temperatures are in good agreement with those from
other recent studies (see Section~\ref{subsec:HaloTandSB}), which used different methods to estimate
the foreground emission. It therefore seems unlikely that a bias in the halo temperature measurements
is causing the discrepancy between the observed temperatures and those predicted by the SN-driven ISM
model. Nevertheless, an accurate model of SWCX emission will help evaluate our method for estimating
the foreground emission.

\section{SUMMARY AND CONCLUSIONS}
\label{sec:Summary}

We have analyzed 26 high-latitude \xmm\ observations of the SXRB, concentrating in particular on the
emission from the Galactic halo. These observations were chosen from a much larger set of
observations (Paper~I) as they are expected to be the least contaminated by SWCX emission. We
modeled the 0.4--5.0~\kev\ X-ray spectra with emission components from the foreground, the Galactic halo, and the
extragalactic background, with additional components modeling parts of the instrumental
background. Assuming a single-temperature CIE plasma model for the halo, we typically obtained halo temperatures
of $\sim$$(\mbox{1.8--2.4}) \times 10^6~\K$, and emission measures of $\sim$0.0005--0.006~\emismeas,
in good agreement with previous studies. While the halo temperature is fairly constant, the emission
measure exhibits significant sightline-to-sightline variation.

We compared the observed X-ray properties of the halo with the predictions of three physical models
for the origin of the hot halo gas: (1) a disk galaxy formation model, which predicts the existence
of a hot halo extended over tens of kiloparsecs \citep{toft02,rasmussen09,crain10}; (2) a model in
which the halo is heated by extraplanar SNe, and the hot gas resides in isolated SNRs
\citep{shelton06}; and (3) a more comprehensive model of SN heating of the ISM, in which the SNRs do
not evolve in isolation, but drive a fountain of hot gas from the disk into the halo \citep{joung06}.

Model~2 matches the observed halo temperature reasonably well, but this model predicts emission at least an order
of magnitude too faint in the \xmm\ band, implying that another source of hot gas is needed, in
addition to isolated extraplanar SNRs. With Model~1, the conclusions are more uncertain: the
original simulations that we examined (from \citealt{rasmussen09}) predicted luminosities at least an order of
magnitude too faint in the \xmm\ band, whereas newer simulations with stronger stellar feedback
predict larger X-ray luminosities \citep{crain10}, in better agreement with those inferred from our
observations. However, the predicted emission from that model includes contributions not only from accreted extragalactic material
falling into the galactic potential well, but also from material that has been heated by stars and flowed out
from the disk. Predictions of the relative X-ray surface brightnesses due to these two processes are
needed to determine whether or not emission from accreted extragalactic material is a major
contributor to the observed halo X-ray emission.

A flow of hot gas from the disk into the halo, in the form of an SN-driven galactic fountain
\citep{shapiro76,bregman80}, also occurs in the SN-driven ISM model (Model~3). The X-ray
surface brightness predicted by this model is in good agreement with the observed surface brightness
of the halo, although it should be noted that the halo in this model has not yet settled down to a
steady state. Therefore, while we cannot currently rule out the possibility of a significant
contribution from accreted extragalactic material, our analysis indicates that hot gas in an
SN-driven galactic fountain is a major (possibly dominant) contributor to the halo X-ray
emission. While previous work has shown that disk SNe produce sufficient energy to heat the halo
\citep[e.g.,][]{wang98,shelton07,yao09,yoshino09}, to the best of our knowledge this is the first
time that CCD-resolution spectra of the halo have been compared with predictions from a hydrodynamical
model of a galactic fountain.

Although the halo X-ray surface brightness predicted by the SN-driven ISM model is in good agreement
with the observations, this model overpredicts the X-ray temperature of the halo. We discussed
various possible reasons for this discrepancy, including the simulation under-resolving the mixing
of hot and cold gas, the simulation omitting important sources of pressure support that would
increase the disk scale height and lower the temperature at $|z| \sim 1$--2~\kpc, the hot halo gas
being underionized, and our potentially underestimating the foreground \OVII\ intensity in the
observed spectra, which would in turn cause us to underestimate the observed halo temperature
(although this final possibility seems unlikely, given the good agreement between our temperature
measurements and those from other recent studies). On
the modeling side, future simulations that have higher resolution in the halo, that include magnetic
fields, and that track the ionization evolution \citep{benjamin01} may help reduce the
discrepancy between the predicted and observed X-ray temperature. A SWCX model that can more
accurately predict the foreground emission will help on the observational side.

We can further test the current and future simulations with additional observations. We plan to
expand our \xmm\ survey (Paper~I) to cover the whole sky. X-ray absorption line measurements can
also be used to test the models. The significant sightline-to-sightline variation in the observed
halo emission shows that, in order to examine halo models in detail, data from as many sightlines
as possible should be used.

\acknowledgements
We thank Jesper Rasmussen for supplying us with X-ray luminosities
and temperatures derived from disk galaxy simulations, and for helpful discussions.
We also thank Rob Crain for making us aware of \citet{crain10}, and for supplying us
with predictions from their simulations.
We thank the anonymous referee, whose detailed technical comments have improved the data
analysis presented in this paper.
We acknowledge use of the R software package \citep{R}.
This research is based on observations obtained with \xmm, an ESA science mission with instruments
and contributions directly funded by ESA Member States and NASA.
The software used in this work was in part developed by the DOE-supported ASCI/Alliance Center for
Astrophysical Thermonuclear Flashes at the University of Chicago. The simulations described in
Section~\ref{subsec:SNDrivenISM} were performed at the Pittsburgh Supercomputing Center and the
National Center for Supercomputing Applications supported by the NSF.
DBH and RLS acknowledge funding from NASA grant NNX08AJ47G, awarded through the Astrophysics Data Analysis Program.
KJK and RLS acknowledge funding from NASA grant NNX09AD13G, awarded through the Astrophysics Theory and Fundamental Physics Program.
MRJ and M-MML acknowledge funding from NASA/SAO grant TM0-11008X.

\bibliography{references}

\end{document}